\documentclass[transmag]{IEEEtran}
\usepackage{latexsym}
\usepackage{hyperref}

\usepackage{cite}
\usepackage{amsmath,amssymb,amsfonts,bm}
\usepackage{algorithm,algorithmic}
\usepackage{graphicx} 
\usepackage{subcaption}
\usepackage{textcomp}
\usepackage{cuted}   
\usepackage[table]{xcolor}

\usepackage{blindtext}
\usepackage{eso-pic}
\usepackage{setspace}
\usepackage{amsthm}

\theoremstyle{break}
\usepackage[font=small]{caption}

\newtheorem{theorem}{Theorem} 
\newtheorem{lemma}{Lemma} 

\DeclareMathOperator*{\argmax}{arg\,max}
\DeclareMathOperator*{\argmin}{arg\,min}

\def\BibTeX{{\rm B\kern-.05em{\sc i\kern-.025em b}\kern-.08em T\kern-.1667em\lower.7ex\hbox{E}\kern-.125emX}}
\begin{document}

\title{
Low-Complexity Resource Allocation for Dense Cellular Vehicle-to-Everything (C-V2X) Communications}

\author{
\thanks{Manuscript received XXX, XX, 2020; revised XXX, XX, 2020.}
Mohammad Hossein Bahonar, Mohammad Javad Omidi, and Halim Yanikomeroglu
\thanks{M. H. Bahonar and M. J. Omidi are with the Department of Electrical and Computer Engineering, Isfahan University of Technology, Isfahan, 84156-83111, Iran. (email: mh.bahonar@ec.iut.ac.ir; omidi@iut.ac.ir)}
\thanks{H. Yanikomeroglu is with the Department of Systems and Computer Engineering, Carleton University, Ottawa, ON K1S5B6, Canada. (e-mail: halim@sce.carleton.ca) }
}

\IEEEtitleabstractindextext{
\begin{abstract}
Vehicular communications are the key enabler of traffic reduction and road safety improvement referred to as cellular vehicle-to-everything (C-V2X) communications.
Considering the numerous transmitting entities in next generation cellular networks, most existing resource allocation algorithms are impractical or non-effective to ensure reliable C-V2X communications which lead to safe intelligent transportation systems.
We study a centralized framework to develop a low-complexity, scalable, and practical resource allocation scheme for dense C-V2X communications.
The NP-hard sum-rate maximization resource allocation problem is formulated as a mixed-integer non-linear non-convex optimization problem considering both cellular vehicular links (CVLs) and non-cellular VLs (NCVLs) quality-of-service (QoS) constraints.
By assuming that multiple NCVLs can simultaneously reuse a single cellular link (CL),
we propose two low-complexity sub-optimal matching-based algorithms in four steps.
The first two steps provide a channel-gain-based CVL priority and CL assignment followed by an innovative scalable min-max channel-gain-based CVL-NCVL matching.
We propose an analytically proven closed-form fast feasibility check theorem as the third step.
The objective function is transformed into a difference of convex (DC) form and the power allocation problem is solved optimally using majorization-minimization (MaMi) method and interior point methods
as the last step.
Numerical results verify that our schemes are scalable and effective for dense C-V2X communications.
The low-complexity and practicality of the proposed schemes for dense cellular networks is also shown.
Furthermore, it is shown that the proposed schemes outperform other methods {\color{black} up to \%6} in terms of overall sum-rate in dense scenarios and have a near optimal performance.
\end{abstract}

}

\maketitle

\section{Introduction}
\IEEEPARstart{V}{ehicular} communications of autonomous vehicles, on the ground or in the air, are necessary for world traffic reduction and improving road safety.
Considering the traffic reduction advantage, it is important to notice that according to a UK report \cite{S075}, the average car is parked \%96 of the time.
Therefore, an autonomous vehicle can be available to a wide range of people and can be used in a wide range of applications which will lead to traffic reduction.
Considering the safety improvement advantage, autonomous vehicles can provide safe trips to people when in-vehicle sensors as well as information received from external sources are utilized effectively \cite{S074}.
{\color{black}
Vehicular communications is a key enabler to autonomous vehicle and intelligent transportation systems (ITSs) that should be investigated and designed effectively.
}

\begin{figure*}
  \centering
    \includegraphics[trim={1.9cm 3.0cm 2.8cm 5.5cm},clip,width=0.9\linewidth]{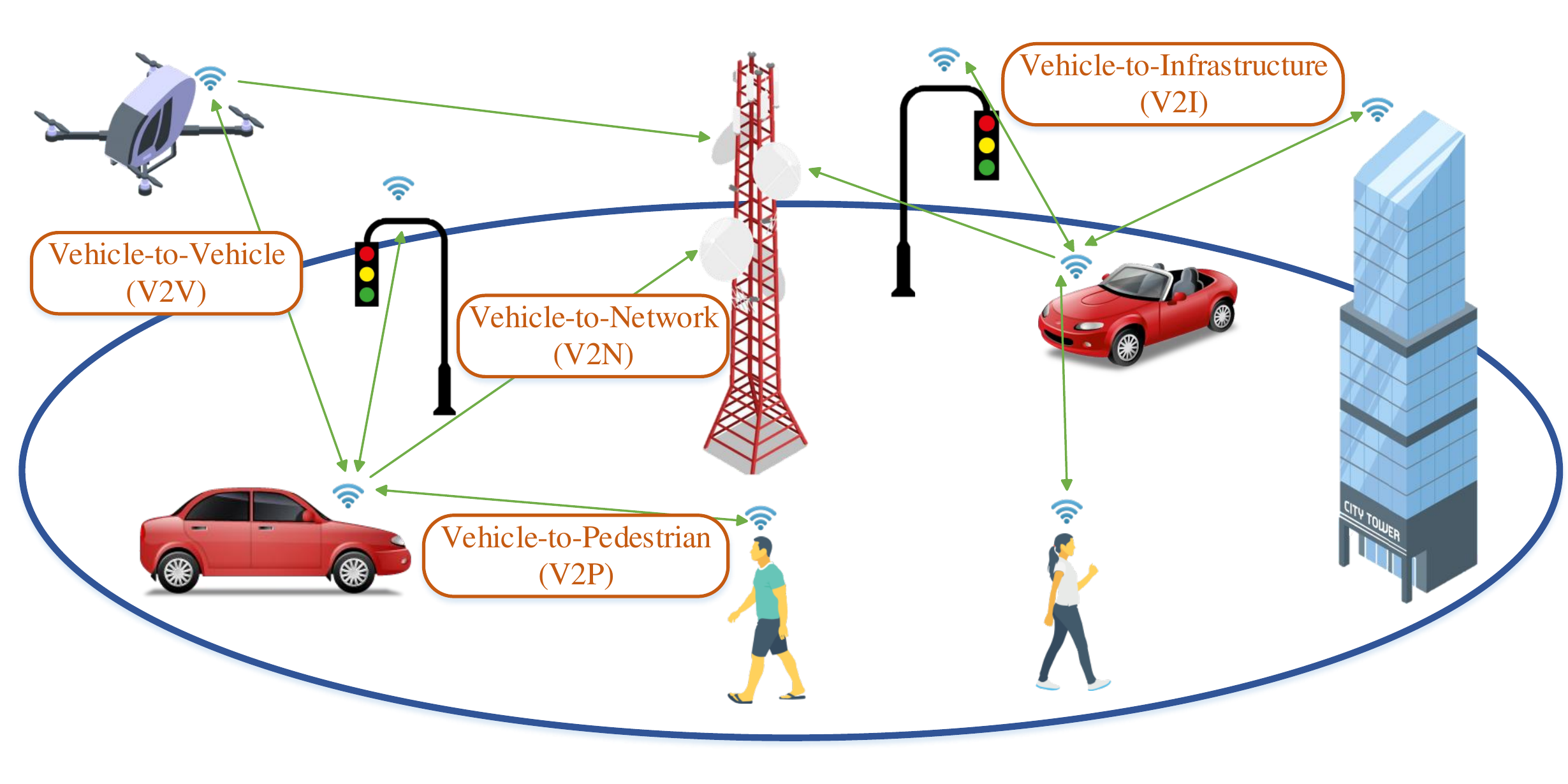}
    \caption{A model of dense C-V2X communications demonstrating vehicle-to-vehicle (V2V), vehicle-to-infrastructure (V2I), vehicle-to-pedestrian (V2P), and vehicle-to-network (V2N) links.}
    \label{FigSysModel}
\end{figure*}

An autonomous vehicle should be able to communicate to other vehicles through vehicle-to-vehicle (V2V) communications that enables data sharing among vehicles without network assistance.
In addition to that and in order to improve the safety aspect of intelligent transportation using autonomous vehicles, it is necessary to consider vehicle-to-network (V2N), vehicle-to-infrastructure (V2I), and vehicle-to-pedestrian (V2P) communications as shown in Fig. \ref{FigSysModel}.
The underlaying technology has been recently gained much attention by academia and industry.
Communications between vehicular user equipments (VUEs) and other transmitting entities are assumed to be assisted by V2N links.
Hence, this technology is referred to as cellular vehicle-to-everything (C-V2X) communications \cite{S076}.
The initial attempts to standardize the technology has been carried out by the third generation partnership project (3GPP) as a part of Release 14.
Due to the important role of this technology in future wireless networks, further development and standardization of C-V2X communications has also been discussed as sidelink enhancement in Release 18.
Basically, C-V2X has been one of the most important use-cases considered in the standardization of 5G, and it
remains to be one of the most important use-cases considered in the pre-standards 6G discussions.
Hence, C-V2X communications is an important aspect of 5G enhancement as well as next generation cellular networks.

It is critical to consider that the number of VUEs as well as the number of transmitting entities will increase in future networks.
Despite the researches that have investigated C-V2X communications, the topic of dense and ultra-dense C-V2X communications has not been investigated thoroughly.
Considering the large number of transmitting entities in dense C-V2X communications and the fact that each VUE should be able to communicate to a sufficient number of transmitting entities to ensure its safety, it is critical to develop resource allocation algorithm for dense C-V2X communications such that the spectral efficiency (SE) of the cell is maximized.
Developing such algorithms allows the VUEs to establish connections to a large number of transmitting entities and to ensure their safety.
To the best of our knowledge, the SE maximization of dense C-V2X has not been investigated in the literature.
Therefore, we study SE maximization of dense C-V2X communications by allocating resource to VUEs and other transmitting entities.
Although C-V2X communications can be implemented on a device-to-device (D2D)-based platform, specific algorithms are required to be designed for dense C-V2X communications that are low-complexity and effective for  future wireless networks.


In order to investigate the mentioned problem, we categorize existing vehicular links (VLs) to cellular VLs (CVLs) that are among VUEs and the base station (BS) and NCVLs that are among VUEs and every other transmitting entities but the BS.
Most of existing researches consider non-dense scenarios where each CL is reused by a limited number of NCVLs, typically at most one NCVL.
In dense networks, the feasibility of the resource allocation problem should be investigated, and it can not be assumed.
Most existing algorithms are not scalable, effective, or feasible for dense C-V2X communications.
In addition, the spatial reuse gain of cells that can be used to maximize the SE of dense networks have not been considered in most existing resource allocation schemes.
Some existing methods do not consider practical assumptions.
In a practical scenario, each NCVL can reuse exactly one CL, both VCLs and NCVLs have minimum quality-of-service (QoS) requirements, and the resource allocation algorithm should have reasonable computational complexity.
Therefore, low-complexity, scalable, feasible, and practical resource allocation algorithms should be developed for dense C-V2X communications in order to maximize the SE of cellular networks in future wireless networks with a large number of transmitting nodes.

{\color{black}
The main aspect of our proposed system model is the dense deployment of transmitting entities in a cell which is an important and challenging aspect of future wireless networks and ITSs.
Additionally, the topic of dense C-V2X has not been investigated previously to the best of our knowledge.
It should be noted that by increasing the cell density, many existing methods become infeasible due to design issues, practicality concerns, or complexity issues while our proposed effective approach remains feasible due to analytically proven necessary and sufficient proposed feasibility check theorem and the scalable NCVL selection method.
In contrast to many existing schemes of D2D communications that add new assumptions to the system model, we consider a general system model where dense deployment of transmitting entities in a C-V2X environment is also assumed.
Therefore, our proposed approach would have more advantages compared to other methods due to its lower complexity and low optimality gap.
}

In this paper, we formulate the centralized uplink resource allocation problem for dense C-V2X communications in future wireless networks.
Dense deployment of NCVLs is the only requirement of the system model which has applications in wireless sensor networks, IoT networks, and next generation cellular networks.
Due to the large number of transmitting nodes in future wireless networks, it is assumed that multiple NCVLs can reuse a single CL.
It is also assumed that each NCVL can reuse a single CL, which is a practical assumption from the point of view of hardware.
We consider a single-cell network since inter-cell interference can be managed efficiently \cite{R042_026_023_017}.
The SE maximization problem is modeled as a mixed-integer non-linear non-convex sum-rate maximization problem with QoS constraints for both VCLs and NCVLs.
The SE maximization problem of users in interference channels has been proven NP-hard in \cite{R005_028}.
We propose two sub-optimal low complexity resource allocation algorithms.
The CVL-NCVL matching is found using a sub-optimal innovative channel-gain-based CVL priority and CL assignment followed by a min-max channel-gain-based NCVL selection.
The matching feasibility is verified using a fast and analytically proven feasibility check theorem.
Considering the difference of convex (DC) form of the objective function and using the Majorization-Minimization (MaMi) technique, the concave-convex procedure (CCCP), and the interior point methods, an iterative optimal power allocation algorithm is developed as the final step.
Our proposed scheme is scalable for the dense C-V2X communications.
Our scalable and practical resource allocation algorithm utilizes the spatial reuse gain of the cell effectively to maximize the SE of the cell using the novel CVL-NCVL matching, the closed-form feasibility check theorem, and the optimal power allocation algorithm.
Numerical results verify that our proposed methods outperform other methods and are efficient for resource allocation to numerous NCVLs related to dense C-V2X communications in future wireless networks.
The main contributions of the paper can be summarized as follows:
\begin{itemize}
\item We propose a channel-gain-based CVL priority assignment followed by an innovative scalable min-max  channel-gain-based NCVL selection to find the CVL-NCVL matching.
{\color{black}
Each new NCVL is selected on the basis of all previously admitted CVLs using the selection approach.
This means that the approach is scalable when the number of admitted NCVLs increases.
}
\item
{\color{black}
We introduce a closed-form and efficient feasibility check theorem which express sufficient and necessary conditions for the feasibility of the resource allocation problem.
The theorem has not been introduced in the literature previously.
The sufficiency and necessity conditions are proven analytically.
The theorem also provides an initial feasible point for overall SE maximization problem.
}
\item We advance two sub-optimal resource allocation algorithms, which consist of a sub-optimal CVL-NCVL matching component and an optimal power allocation scheme, that outperform other methods in dense scenarios.
\item The spatial reuse gain of the cell is utilized effectively by considering the multiple reuse assumption of CLs  by NCVLs and the innovative CVL-NCVL matching.
\item
{\color{black}
The complexity of our proposed scheme is lower than that of other schemes since we propose to utilize the innovative and scalable CVL-NCVL matching and the closed-form and analytically proven feasibility check theorem.
}
\item
{\color{black}
By comparing the performance of our approach to the optimal approach, it can be verified that our proposed scheme has a near optimal sum-rate performance.
}
\end{itemize}

The rest of the paper is organized as follows.
Related works are summarized in Section \ref{SecRelatedWork}.
In Section \ref{SecSysModel}, we describe the system model and formulate the resource allocation problem.
The proposed four-step centralized scheme is presented in Section \ref{ResAll}.
In Section \ref{SecNumeric} numerical results are reported.
Section \ref{Conc} concludes the paper.

\section{Related Works}
\label{SecRelatedWork}
In this section, we review recent researches of the literature.
The topics of V2X and C-V2X communications, uplink and downlink spectrum reuse, resource allocation to a single CVL or NCVL, resource allocation to multiple CVLs or NCVLs, dense networks, and resource allocation to dense C-V2X communications are investigated in the following paragraphs.
Since the topics of V2X and C-V2X communications are closely related to the topic of D2D communications, D2D communications related researches are also reviewed in this section in order to present a comprehensive vision of the literature.

V2X communications in the unlicensed band have limited usage due to their short-range coverage and the fact that QoS requirements of all links may not always be guaranteed \cite{S067}.
The reason is that the data transmissions related to NCVLs are not controlled by a BS when the unlicensed spectrum is utilized.
In addition to that, the interference may not be addressed effectively, since a wide range of users and devices may use the unlicensed spectrum.
Therefore, cellular spectrum is proposed in the 3GPP Release-18 Sidelink enhancement where C-V2X communications is a big topic \cite{S068}.
Despite these challenges, we study V2X communications in cellular systems denoted as C-V2X for their potential to improve SE.
One of the main challenges of C-V2X communications is the intra-cell interference caused by NCVLs to CVLs.
In the overlay scheme of C-V2X communications, dedicated cellular resources are assigned to NCVLs.
Thus, there is no interference between NCVLs and CVLs.
In the underlay scheme, however, NCVLs and CVLs use the same spectrum, and interference management is critical.
The spectrum that is assigned to CVLs in the overlay scheme is not effectively utilized since the spatial reuse gain is not considered.
As a result, the underlay scheme can improve the SE of the cell more than the overlay scheme.
Further, uplink spectrum reuse can provide greater SE compared to the downlink spectrum due to the underutilization of the latter.
In this paper, we study the uplink C-V2X communications underlaying dense cellular networks where interference coordination and resource allocation are two of the main challenges.

The uplink resource allocation problem can be solved in a centralized or distributed manner.
Centralized approaches are usually more complex and effective compared to distributed approaches.
Game theory is an effective tool for designing distributed resource allocation schemes  \cite{R151,R077_2} for D2D communications as well as C-V2X communications, while optimization approaches are usually utilized in centralized schemes \cite{R009}.
The authors of and \cite{R151} used auction theory and Stackelberg game modeling to propose distributed resource allocation algorithms, respectively.
The authors of \cite{R151} proposed a benchmark for system performance and compared the performance of their proposed system with the benchmark.
The benchmark is a centralized resource allocation scheme that was developed using convex optimization techniques.
The authors of \cite{R077_3} used game theory as a tool for distributed resource allocation in dense cellullar networks.
Due to the higher performance of centralized approaches, we aim to design a centralized resource allocation scheme.

A single CVL or NCVL might be considered in a resource allocation or performance evaluation problem.
Cellular UEs (CUEs)  and D2D pairs in D2D communications can be viewed equivalent to CVLs and NCVLs in C-V2X communications.
The { authors of \cite{R142} investigated cooperative communication considering} one CUE and one D2D pair in the cell {and computed the outage probability of the CUE and the average data rate of the D2D pair.}
{The authors of \cite{R009_006} introduced the concept of an interference-limited area (ILA) corresponding to the area of a cell that receives little interference.
The ILA concept was used in \cite{R009_006} to enhance the capacity of D2D communications.}
In \cite{R140_009}, { a capacity oriented algorithm was introduced to allocate resources to a single D2D pair} reusing CLs of multiple CUEs.
Since just one NCVL was considered, the proposed methods of \cite{R142,R009_006,R140_009} resulted in a lower SE improvement than works with multiple NCVLs.
Hence, the researches assuming a single NCVL or CVL are not suitable for dense C-V2X communications in future wireless networks.

Multiple D2D pairs assumption which is equivalent to multiple NCVLs was investigated in \cite{R003,R042_026_020,R050,R136} where at most one D2D pair could reuse each CL and each CL could be shared with at most one D2D pair.
The authors in \cite{R003} allocated optimal power to a fully loaded system with an  equal number of CUEs and D2D pairs using the maximization on the boundary property of the objective function followed by the Kuhn-Munkres algorithm \cite{R003_016}.
This approach is suitable for non-dense C-V2X communications since it is necessary to assume that the number of NCVLs is  equal to the number of CVLs.
In \cite{R042_026_020}, resource allocation for relay-aided D2D communications  involving channel uncertainty was investigated, and a distributed solution for sum-rate maximization using a gradient aided dual decomposition algorithm was proposed.
The authors of \cite{S072} also investigated the channel uncertainty problem for beyond 5G C-V2X communications.
However, the authors maximized the energy efficiency (EE) of the network in a non-dense environment.
In \cite{R136}, a joint mode selection and resource group assignment algorithm with polynomial time {was} proposed for relay-aided D2D communications.
The authors of \cite{R003,R042_026_020,R050,R136} assumed that multiple NCVLs could reuse the whole spectrum, but each CL could be used by at most one NCVL.
Depending on the spatial reuse gain of the cell, multiple NCVLs can reuse the same CL simultaneously if interference is addressed properly.
Thus, the spatial reuse gain of the cell was not utilized in \cite{R003,R042_026_020,R050,R136}.
Resource allocation for V2X communications using deep neural networks was addressed in  \cite{S071, S073}.
The authors of  \cite{S071} considered a limited number of transmitting entities and only V2I links.
The authors of \cite{S073} focused on mobile edge computing in non-dense environments.
Considering the large number of transmitting nodes in future wireless networks, taking advantage of the spatial reuse gain of the cell is critical.

To meet the rate requirements of future wireless networks with a large number of transmitting nodes, dense C-V2X communications can be utilized if resource allocation algorithms are developed properly.
In order to develop practical resource allocation algorithms for dense C-V2X communications in future cellular networks, the feasibility of resource allocation problem, its practicality, and its complexity should be considered.
In addition to that, the spatial reuse gain of cells should be utilized effectively.
Therefore, it is necessary to assume that multiple NCVLs can reuse one CL at the same time, which is a more challenging assumption and can result in a dramatically higher SE.
Since C-V2X communications can be implemented on a D2D platform, it is necessary to consider proposed schemes in the field of D2D communications.
Some researchers have assumed that one D2D pair may use multiple CLs \cite{R135,R141} which is not a practical assumption from a hardware point of view, as each D2D pair may need multiple transmitter modules.
Therefore, these approaches are not suitable practical C-V2X communications.
A sub-optimal graph-coloring based algorithm where D2D pairs are viewed as a set of vertexes was proposed in \cite{R135}.
However, the approach was heuristic and focused on the downlink spectrum.
The system model of \cite{R135} was not practical since the usage of multiple CLs by one D2D pair was assumed by the authors.
A resource allocation algorithm consisting of a channel assignment phase followed by a reuse phase was developed in \cite{R141}.
To admit a D2D pair, a minimum QoS requirement is assumed for previously admitted D2D pairs during the reuse phase, which prevents resource allocation to { a} large number of D2D pairs in a dense cellular network.
Thus, the proposed scheme is not suitable for dense C-V2X communications.
Distributed and game-theoretical approaches were investigated in \cite{R011,R042_026_025}. 
The authors of \cite{R011} addressed the interference management problem properly using a pricing-based resource sharing algorithm where interference costs were defined for all CLs.
The proposed method did not result in SE maximization but tried to guarantee QoS requirements of CUEs.
A large value for the distance between D2D pairs was also assumed, which was not a practical assumption.
The authors of \cite{R042_026_025} proposed a coalitional game based scheme for the resource allocation problem.
The convergence of the proposed scheme to the Nash-stable equilibrium was also proven.
The authors of \cite{S078}, investigated power allocation problem in drone-assisted V2X communications.
However, the main focus of the paper was on trajectory design and the usage of a unmanned-aerial-vehicle as a relay, while resource allocation to dense C-V2X communications in future wireless networks was not considered.
The authors of \cite{S069}, proposed a V2V resource allocation based on C-V2X communications.
However, latency reduction was the main focus of the paper and dense C-V2X communications was not considered.
The authors of \cite{S070}, investigated the resource allocation problem for V2X communications by formulating the problem as a three dimensional matching problem.
However, the sum-rate of maximization of all transmitting entities was not considered and a local search based approximation algorithm was utilized.
Considering the hardware limitations of D2D pairs, it is more reasonable to assume that each D2D pair can reuse just one CL \cite{R077}.
The maximization on the boundary property of the objective function was used in \cite{R077}.
The authors of \cite{R042_026} modeled the resource allocation problem as a mixed-integer programming problem and {solved} it using a subchannel sharing protocol.
The resource sharing possibility among each two D2D pairs was investigated in the proposed subchannel sharing protocol.
However, only the QoS of CUEs was considered, and the sum-rate of D2D pairs was maximized.
Additionally, the approach was not scalable, and therefore not practical for ultra-dense scenarios.


\section{System Model and Problem Formulation}
\label{SecSysModel}
{ To facilitate understanding, we list frequently used notations in Table \ref{tableSymNotations}.}

\subsection{System Model and Assumptions}
\label{SubSecSysModel}
{\color{black}
We present a vehicular system model that can be used in ITSs and autonomous driving.
Increasing the capacity of such a network can be used to enable higher data exchange between transmitting entities using V2V, V2N, V2P, or V2P links.
This data exchange is a key enabler for ITS and can be used to in congestion control.
This higher data rate can also be used to improve road safety and decrease travel times.
}

We consider the uplink spectrum of a single macro-cell dense cellular network with $N$ equal bandwidth and orthogonal CLs.
As illustrated in Fig. \ref{FigSysModel}, there exist $N$ vehicular UEs (VUEs) in the cell that communicate with the BS using V2N links.
These VUEs use cellular resources and their links are denoted as CVLs.
In addition to VUEs there exist other transmitting entities that communicate with the VUEs such as  smart traffic lights, pedestrian, and other VUEs that use V2I, V2P, and V2V communication links, respectively.
We assume that these transmitting entities reuse the available CLs instead of utilizing dedicated cellular resources and their links are denoted as NCVLs.
Since the number of available CLs are equal to the number of VUEs and CVLs, the cellular system is a fully loaded system.
The set of CVLs amongs VUEs and the BS is denoted by $\bm{\mathcal{C}} = \{c_1, c_2, ..., c_N\}$.
The set of NCVLs among VUEs and other transmitting entities is denoted by $\bm{\mathcal{D}} = \{d_1, d_2, ..., d_M\}$.
The number of CVLs and NCVLs is equal to $N$ and $M$, respectively.
In order to have practical assumption and due to the fact that each VUEs might communicate with multiple transmitting entities through NCVLs, it can be concluded that $M>N$.
The transmitter(TX) and receiver(RX) of each CVL are one of VUEs and BS, respectively.
The TX and RX of each NCVL are one of the VUEs and one of the transmitting entities, respectively, thar are placed in proximity to each other.
Each CL is used by exactly one CVL and each CVL occupies exactly one CL.
Since the TX module of each transmitting entity can transmit at just one CL at a time, it is reasonable to assume that each NCVL can reuse the CL assigned to one NCVL, while multiple CVLs can share the same CL simultaneously.

\begingroup
\setlength{\tabcolsep}{6pt} 
\renewcommand{\arraystretch}{1.3}
\begin{table*}[t]
\begin{center}
\caption{Symbol Notations}
\label{tableSymNotations}
\begin{tabular}{|>{}p{1cm}|>{}p{7cm}||>{}p{1cm}|>{}p{7cm}|}
\hline
Symbol & Description & Symbol & Description \\ \hline
     $N$   &  The number of CVLs       	   &  $M$  & The number of NCVLs   \\ \hline
     $\mathcal{C}$   &  The set of CVLs      & $\mathcal{D}$  & The set of NCVLs \\ \hline
 $h^c_i$ & The channel gain from the TX $c_i$ to the BS &  $h^d_j$ & The channel gain from the TX of $d_j$ to its RX \\ \hline
$h^{d,b}_j$ & The channel gain from the TX of $d_j$ to the BS & $h^{c,d}_{i,j}$ & The channel gain from the TX of $c_i$ to the RX of $d_j$ \\ \hline
$h^{d,d}_{i,j}$ & The channel gain from the TX of $d_i$ to the RX of $d_j$ & $N_i$ & The number of NCVLs pairs reusing  $c_i$ \\ \hline
$r^c_i$ & The spectral efficiency of CVL $c_i$ & $r^d_j$ & The spectral efficiency of NCVL $d_j$ \\ \hline
$r^{c,\rm{min}}_i$ & The minimum QoS requirement of CVL $c_i$ in terms of spectral efficiency &
$r^{d,\rm{min}}_j$ & The minimum QoS requirement of NCVL $d_j$ in terms of spectral efficiency \\ \hline
$p^c_i$ & The transmission power of the TX of $c_i$ & $p^d_j$ & The transmission power of the TX of $d_j$ \\ \hline
$\psi_{i,j}$ & The CVL-NCVL resource sharing indicator of $c_i$ and $d_j$ & $\rho_j$ & The NCVL resource sharing indicator of $d_j$ \\ \hline
$\bm{{\color{black}p}}$ & The transmission power vector of the TXs of all CVLs and all NCVLs & $\bm{\Psi}$ & The CVL-NCVL resource sharing matrix of all CVLs and all NCVLs\\ \hline
$P^{c,\rm{max}}_i$ & The maximum transmission power of the TX of $c_i$ & $P^{d,\rm{max}}_j$ & The maximum transmission power of the TX of $d_j$ \\ \hline
$\alpha_1$ & The index of the CVL with highest priority & $\alpha_N$ & The index of the CVL with lowest priority \\ \hline
$\bm{\beta}_i$ & The set of NCVL indices reusing $c_i$ &
$\beta_{i,k}$ & The index of the $k$th NCVL added to $\bm{\beta}_i$ \\ \hline
$g^c_i$ & The channel gain from the TX of $c_{\alpha_i}$ to the BS & $g^{d,b}_j$ & The channel from the TX of $d_{\beta_{i,j}}$ to the BS \\ \hline
$g^d_j$ & The channel gain from the TX of $d_{\beta_{i,j}}$ to its RX & $g^{c,d}_{i,j}$ & The channel gain from the TX of $c_{\alpha_i}$ to the RX of $d_{\beta_{i,j}}$ \\ \hline
$g^{d,d}_{k,j}$ &The channel gain from the TX of $d_{\beta_{i,k}}$ to the RX of $d_{\beta_{i,j}}$ & $\bm{{\color{black}p}}_i$ & The transmission power vector of the TX of $c_{\alpha_i}$ and the TX of all NCVLs that are reusing $c_{\alpha_i}$ \\ \hline
$\bm{{\color{black}p}}_i^{\rm{max}}$ & The maximum transmission power vector of the TX of $c_{\alpha_i}$ and the TXs of NCVLs  that are reusing $c_{\alpha_i}$  & $\bm{H}_i$ & The channel gain matrix among the TX of $c_{\alpha_i}$ and the RXs of NCVLs that are reusing its $c_{\alpha_i}$ \\ \hline
$R^i(.)$ & The sum-rate function of $c_{\alpha_i}$ and the NCVLs using its CL &  $h^{c,(l_i)}_i$ & The channel gain from $c_i$ to the BS using the $l_i$th CL  \\ \hline
$R_{vex}^i (.)$ & The convex part of $R^i(.)$ & $R_{cav}^i (.)$ & The concave part of $R^i(.)$\\ \hline
\end{tabular}
\end{center}
\end{table*}
\endgroup

The BS is located at the center of the macro-cell and has greater maximum transmission power capability than other UEs.
For this reason, the downlink interference caused by the BS exists at most parts of the cell and all CLs in a fully loaded scenario.
However, the uplink interference mostly exists around the VUEs that use the corresponding CL.
So the uplink spectrum is less utilized \cite{R003_009} and more suitable for C-V2X communications.
We assume half-duplex communication among the TX and RX of any NCVLs.
It should be noted that a full-duplex communication can be modeled as two half-duplex communications
{\color{black} with different cellular links where the TX (RX) of the first (second) communication link is also the RX (TX) of the second (first) one.
It should be noted that the communication among VUEs and the BS happens in the uplink spectrum where VUEs transmit data to the BS.
}


Due to the fully loaded assumption, there exist $N$ CLs denoted by $\mathcal{N}= \{n_1, n_2, ..., _N \}$.
Each CVL and NCVL transmit data at one of the CLs. Hence, the channel gains depend on the CL assignment.
Without loss of generality, for notational simplicity, and to increase the readability of the proposed method, we assume that the following introduction of channel gain values among CVLs, NCVLs, and the BS corresponds to a situation where the CL assignment is performed for all CVLs and NCVLs.
Therefore, it would be sufficient to propose a CL assignment prior to the power allocation procedure.
We propose a CL assignment procedure to CVLs and NCVLs alongside our proposed CVL priority assignment and NCVL selection procedure in Section \ref{SubCUEPri} and Section \ref{SubD2DMatch}, respectively.
Therefore, the CL assignment is considered and after the CL assignment channel gains can be introduced.
The desired link channel gains from the TX of $c_i$ to the BS and from the TX of $d_j$ to its corresponding RX after the CL assignment are denoted by $h^c_i$ and $h^d_j$, respectively.
The channel gains of the interference links between the TX of $d_j$ and the BS, between the TX of $c_i$ {and} the RX of $d_j$, and between the TX of $d_j$ {and} the RX of $d_k$ after the CL assignment are denoted by $h^{d,b}_j$, $h^{c,d}_{i,j}$, and $h^{d,d}_{j,k}$, respectively.
The additive white Gaussian noise (AWGN) power on each CL is denoted by $\sigma^2$.
Minimum QoS requirements of $c_i$ and $d_j$ that are requested from the BS
in terms of spectral efficiency are denoted by $R_{i}^{c,\rm{min}}$ and $R_{j}^{d,\rm{min}}$, respectively.

It should be noted that in case of imperfect channel state information (CSI), channel gains can be approximated with distance which is a different research topic.
{\color{black}
The distance can also be estimated using geospatial data, artificial-intelligence/machine-learning-based techniques, or measurement reports of long-term evolution (LTE) cellular systems.}


\subsection{Spectral Efficiency and  Data Rate}
Spectral efficiencies of $c_i$ and $d_j$ are denoted by $R^c_i \triangleq \log_{2}(1+\gamma^c_i)$ and $R^d_j \triangleq  \log_{2}(1+\gamma^d_j)$, respectively.
The received signal to interference and noise ratios (SINRs) of $c_i$ and $d_j$ are also denoted by $\gamma^c_i$ and $\gamma^d_j$, respectively.
The SINRs $\gamma_i^c$ and $\gamma_j^d$ can be expressed as
\begin{align}
\label{CUESINREq}
\gamma^c_i &= \frac{p^c_i h^c_i}{\sigma^2
+\sum\limits_{j=1}^{M} \psi_{i,j} p^d_j h^{d,b}_j}, \\
\label{D2DSINREq}
\gamma^d_j &= \frac{p^d_j h^d_j}{\sigma^2
+\sum\limits_{i=1}^{N} \psi_{i,j} p^c_i h^{c,d}_{i,j} + \sum\limits_{k=1 \atop k\neq j}^{M} \sum\limits_{i=1}^{N}
\psi_{i,j} \psi_{i,k} p^d_k h^{d,d}_{k,j}},
\end{align}
respectively,
where $p^c_i$ and $p^d_j$ designate the transmission powers of the TX of $c_i$ and the TX of $d_j$, respectively.
The CVL-NCVL resource sharing indicator variable is denoted by $\psi_{i,j}$,
$\psi_{i,j}=1$ when $c_i$ and $d_j$ are the same CL; otherwise, $\psi_{i,j}=0$.
The data rate of each CVL and NCVL is equivalent to the multiplication of its spectral efficiency and the CL bandwidth.

\subsection{Resource Allocation Problem Formulation}
\label{ProbFormulation}

The objective of our centralized resource (i.e., CLs and transmission powers) allocation problem is to jointly assign the CLs and transmission powers to all CVLs and NCVLs that are operating in an underlaying manner to maximize the overall sum-rate of the cell.
Since $M>N$, multiple NCVLs may reuse the same CL.
A NCVL that is able to reuse a CL is called an link.
Due to the CL bandwidth equality assumption, the sum-rate maximization problem is equivalent to the spectral efficiency summation maximization of all CVLs and NCVLs.
As a result, the objective function of the sum-rate maximization problem can be expressed as follows:
\begin{align}
\label{Eq_Rt}
&R^T = \sum_{i=1}^N R^c_i + \sum_{j=1}^M \rho_j R^d_j,\\
\label{Eq_Rt2}
&\rho_j = \sum^N_{i=1} \psi_{i,j}, ~~ j=1,...,M,\\
\label{PsiDef}
&\psi_{i,j} \in \{0,1\}, ~~\forall i=1,...,N,~~ j=1,...,M,
\end{align}
where $\rho_j$ is the NCVL resource sharing indicator variable,
$\rho_j=1$ when $d_j$ reuses any CVL;  otherwise, $\rho_j=0$.
It should also be noted that due to the dense deployment of NCVLs and the fact that each NCVL and CVL requests a minimum QoS requirement from the BS, allocating resources to all NCVL may not be possible.
Due to this, some NCVL may not be able to establish a connection between their TX and RX, which is the main reason for the definition of the NCVL resource sharing indicator variable.

We define $\bm{{\color{black}p}} \triangleq {\color{black}[} (\bm{p}^c)^T, (\bm{p}^d)^T {\color{black}]}^T$,
 where
 $\bm{p}^c \triangleq {\color{black}[} p^c_1, p^c_2, ..., p^c_N {\color{black}]}^T $
and
$\bm{p}^d \triangleq {\color{black}[} p^d_1, p^d_2, ..., p^d_M {\color{black}]}^T $
are the transmission power vectors of the TXs of all CVLs and NCVLs, respectively.
We also define the CVL-NCVL resource sharing matrix as
$\bm{\Psi} \triangleq {\color{black}[} \bm{\psi}_1, \bm{\psi}_2, ..., \bm{\psi}_M  {\color{black}]}$, where $\bm{\psi}_j \triangleq {\color{black}[} \psi_{j,1}, \psi_{j,2}, ..., \psi_{j,N}  {\color{black}]}^T$.
The minimum QoS requirements of $c_i$ and $d_j$ can also be expressed in terms of their SINRs as
$\gamma^{c,\rm{min}}_i = 2^{ R^{c,\rm{min}}_i } -1$ and
$\gamma^{d,\rm{min}}_j = 2^{ R^{d,\rm{min}}_j } -1$, respectively.
The overall optimization problem can be formulated as the following mixed-integer non-linear non-convex optimization {problem:}
\begin{subequations}
\label{OptProbEq}
\begin{alignat}{4}
\nonumber
&\max_{\bm{P},\bm{\Psi}} ~~&&R^T,\\
\nonumber
& \rm{s.t.} &&\eqref{PsiDef},\\
\label{QosCUECon}
& && \gamma_i^c \geq \gamma_i^{c,\rm{min}}, ~~&&& i=1,...,N,\\
\label{QosD2DCon}
& && \gamma_j^d \geq \rho_j \gamma_j^{d,\rm{min}}, ~~&&& j=1,...,M,\\
\label{MaxPcCon}
& && 0 \leq p^c_i \leq P^{c,\rm{max}}_i, ~~&&& i=1,...,N,\\
\label{MaxPdCon}
& && 0 \leq p^d_j \leq P^{d,\rm{max}}_j, ~~&&& j=1,...,M,\\
\label{MaxD2DPerCUE}
& &&\rho_j \in \{ 0, 1 \}, ~~&&& j=1,...,M,
\end{alignat}
\end{subequations}
where the maximum transmission powers of the TX of $c_i$ and the TX of $d_j$ are denoted by $P^{c,\rm{max}}_i$ and $P^{d,\rm{max}}_j$, respectively.
Constraints \eqref{QosCUECon} and \eqref{QosD2DCon} are the minimum QoS constraints of CVLs and NCVLs, respectively.
Constraints \eqref{MaxPcCon} and \eqref{MaxPdCon} guarantee that the transmission powers of CVLs and NCVLs are positive and less than the maximum limit.
Constraint \eqref{MaxD2DPerCUE} ensures that each NCVL can reuse at most one CL.
Considering \eqref{Eq_Rt2}, constraint \eqref{MaxD2DPerCUE} implies that $\bm{\Psi}$ has only one {``1"} element in each column and may have several {``1"} elements in each row.

\section{Proposed Resource Allocation Algorithm}
\label{ResAll}
In this section, we propose two low complexity sub-optimal algorithms for optimization problem \eqref{OptProbEq}, consisting of a sub-optimal matching algorithm and an optimal power allocation algorithm.
We aim to maximize the overall sum-rate of the cell in a centralized manner.
Our proposed algorithms have four steps that are introduced in the sub-sections below.

In an optimal scheme, due to the cell density assumption ($M>N$), multiple admitted NVCLs may reuse a single CL, and some NCVLs may remain unadmitted.
The optimal solution is to compute the sum-rate of every possible CVL-NCVL matching using optimal power allocation with exponentially growing complexity with respect to $M$, but this is not practical.

Since the UEs that are using a specific CL do not cause interference to the UEs that are using other CLs, we propose solving the resource allocation problem in a CL based manner where CVLs are prioritized at the first step.
While each CL corresponds to a specific CVL, multiple NCVLs are selected to use the CL at the second step.
Steps 1 and 2 result in an innovative sub-optimal CVL-NCVL matching algorithm with linear complexity with respect to $M$.
In order to verify the CVL-NCVL matching feasibility, the third step involves a fast, low complexity, and novel mathematical check method.
An optimal power allocation method derived from the DC form of the objective function and the MaMi method is the final step.
Different combinations of these steps are introduced as the two proposed resource allocation algorithms.

\subsection{CVL Priority and CL Assignment (Step 1)}
\label{SubCUEPri}
We propose to assign higher priority to the CVLs that can share their CL with a greater number of NCVLs during the matching process.
Due to the dense cell assumption, this approach takes advantage of the spatial reuse gain of the cell and can result in a higher overall sum-rate.
Sharing the CL of a CVL with a greater number of NCVLs depends on the area of the cell where the TX of the CVL can not cause a significant amount of interference to that area.
Inspired by \cite{R009_006}, this area can be denoted as the ILA of the CVL.
Hence, a higher priority should be assigned to a CVL with a greater ILA so that the CVL will be able to share its CL with a greater number of NCVL.
A lower channel gain from the TX of each CVL to BS corresponds to a case where a greater area of the cell receives little interference or equivalently a greater ILA.
Therefore, we propose using the channel gain between the TX of each CVL and the BS as a CVL priority assignment metric.

While each CVL is equivalent to exactly one CL, the CL should be selected among $N$ available ones.
Therefore, we propose assigning CLs to CVLs jointly with the priority assignment procedure.
Denoting the channel gain from the TX of $c_i$ to the BS using the $l_i$th CL as $h^{c,(l_i)}_i$, the CVL priority and CL assignment procedure can be summarized as Algorithm \ref{AlgCUECL}.
The CVL with the $k$th highest priority and its CL are selected as follows:
\begin{equation}
\label{EqJointCUEPriorityCL}
{(\alpha_k, l_k^*)} = \underset{i \in \bm{U}^C, l_i \in \bm{U}^N}{\argmax} h^{c,(l_i)}_i,
\end{equation}
where $\alpha_k$ and $l_k^*$ are the indices of the CVL with the $k$th highest priority and its corresponding CL, respectively.
The index set of available CLs and the index set of CVLs where a CL has not been assigned to them are denoted as
$\bm{U}^N$ and { $\bm{U}^C$}, respectively.
The set formed by elements of $A$ that are not in $B$ is expressed using the set difference operation as $A \backslash B$.
After the CL assignment, each CL can be recognized by its corresponding CVL index since the cell is fully loaded and each CVL corresponds to exactly one CL.
Thus, the superscript representing the CL index can be removed from the channel gain definition as formulated in the fifth line of Algorithm \ref{AlgCUECL} and as described in Section \ref{SubSecSysModel}.
The CVL with the highest (lowest) priority and its corresponding CL are found at the first (last) iteration of Algorithm \ref{AlgCUECL}.
\begin{algorithm}[t]
\caption{CVL priority and CL assignment}
\label{AlgCUECL}
{
\begin{algorithmic}[1]
\STATE $\bm{U}^N = \{1, 2, ..., N \}$, $\bm{U}^C = \{1, 2, ..., N \}$
\STATE Define CVL priority set $\bm{\alpha} = \{ \alpha_1, \alpha_2, ..., \alpha_N \}$
\FOR {$k = \{1, 2, ..., N\}$}
\STATE Compute ${(\alpha_k, l_k^*)}$ from \eqref{EqJointCUEPriorityCL}
\STATE $h^c_{\alpha_i} = h^{c,(l_i^*)}_{\alpha_i} $
\STATE $\bm{U}^C = \bm{U}^C \backslash {\{} \alpha_k {\}}$
\STATE $\bm{U}^N = \bm{U}^N { \backslash \{ } l_k^* {\}}$
\ENDFOR
\end{algorithmic}
}
\end{algorithm}
Therefore, we conclude that
\begin{equation}
\label{DistEQ_Ch}
h^{c}_{\alpha_1} < ... < h^{c}_{\alpha_N},
\end{equation}
where $c_{\alpha_1}$ ($c_{\alpha_N}$)
denotes the CVL with highest (lowest) priority and $h^c_{\alpha_1}$ ($h^c_{\alpha_N}$) denotes the channel gain from the TX of $c_{\alpha_1}$ ($c_{\alpha_N}$) to the BS after the CL assignment.

\subsection{NCVL Selection (Step 2)}
\label{SubD2DMatch}
After the selection of a CVL and its corresponding CL, unadmitted NCVL can be matched to the CVL.
One NCVL is selected at each iteration of this step using our proposed innovative min-max  channel-gain-based approach.
We define the index set of the NCVLs that reuse $c_{\alpha_i}$ as $\bm{\beta}_i \triangleq \{ \beta_{i,1}, ..., \beta_{i,N_i}\}$, where ${\beta}_{i,k}$ is the index of the $k$th NCVL that has been added to the set and $N_i$ is the number of NCVLs that are matched with $c_{\alpha_i}$.
Thus, the index of the selected NCVLs is added to $\bm{\beta}_i$ as discussed below.
The algorithmic representation of the approach is also presented in Algorithm \ref{D2DSelAlg}.
It should be noted that due to the fully loaded assumption of the system model, each CL is recognized with exactly one CVL.
For this reason, the CL assignment of NCVLs is also addressed in our proposed NCVL selection procedure.

An unadmitted NCVL that causes little interference to the TXs of $\bm{\beta}_i$ and receives little interference from their RXs should be chosen as the next selected NCVL.
We propose selecting the unadmitted NCVL $k^*$ as the next admitted NCVL such that its maximum channel gain from the TXs of $c_{\alpha_i}$ and NCVL of $\bm{\beta}_i$ is minimized.
The NCVL selection approach is formulated using a min-max channel-gain-based approach as follows:
\begin{subequations}
\label{SelNextD2DCh}
\begin{align}
\label{MinMax20}
k^* &= \argmin_{k \in \bm{U}} ( \max( s_{k_1}, s_{k_2} )),\\
\label{MinMax7}
s_{k_1} &= \max( h^{c,d}_{\alpha_i,k}, h^{d,d}_{\beta_{i,1},k}, ... , h^{d,d}_{\beta_{i,N_i},k}), \\
\label{MinMax8}
s_{k_2} &= \max(h^{d,b}_k, h^{d,d}_{k,\beta_{i,1}}, ... , h^{d,d}_{k,\beta_{i,N_i}}),
\end{align}
\end{subequations}
where $\bm{U}$ {denotes} the set of unadmitted NCVL.
The channel gain corresponding to the worst-case interference received from (caused to) the TXs (RXs) of $c_{\alpha_i}$ and $\bm{\beta}_i$ is denoted by $s_{k_1}$ ($s_{k_2}$).
The new NCVL is selected with \eqref{SelNextD2DCh}, which considers all previously admitted NCVLs of the corresponding CL.
Therefore, the method is scalable when the number of admitted NCVLs increases, which makes the approach practical for the numerous transmitting nodes and links in future wireless networks and dense cellular environments.

Due to the proposed CVL-NCVL matching procedure, it is guaranteed that $\psi_{\alpha_i,{\beta_{i,j}}} = 1$ and  $\rho_{\beta_{i,j}}=1, \forall j=1,...,N_i$.
Thus, the binary resource sharing indicators can be removed from \eqref{OptProbEq}.
So, the matching procedure converts the sum-rate maximization problem \eqref{OptProbEq} to {$N$} parallel optimizations corresponding to $N$ CLs.
Considering $c_{\alpha_i}$, the optimization problem of this CL can be expressed as follows:
\begin{subequations}
\label{EqOptPrb2}
\begin{alignat}{3}
\label{EqOpt112}
&\max_{\bm{P}_i } \quad &&
{
R^i (\bm{{\color{black}p}}_i) \triangleq R^c_{\alpha_i} (\bm{{\color{black}p}}_i) + \sum_{j=1}^{N_i} R^d_{{\beta}_{i,j}} (\bm{{\color{black}p}}_i),} \\
\label{EqOpt122}
&\rm{s.t.} && \gamma^c_{\alpha_i} \geq \gamma_{\alpha_i}^{c,\rm{min}}, \\
\label{EqOpt132}
& && \gamma^d_{\beta_{i,j}} \geq \gamma_{\beta_{i,j}}^{d,\rm{min}},  j=1,2,...,N_i, \\
\label{EqOpt152}
& && 0 \leq p^c_{\alpha_i} \leq P^{c,\rm{max}}_{\alpha_i}, \\
\label{EqOpt162}
& && 0 \leq p^d_{\beta_{i,j}} \leq P^{d,\rm{max}}_{\beta_{i,j}} ,  j=1,2,...,N_i,
\end{alignat}
\end{subequations}
where
$\bm{{\color{black}p}}_i \triangleq {\color{black}[} p^c_{\alpha_i}, p^d_{\beta_{i,1}}, p^d_{\beta_{i,2}}, ..., p^d_{\beta_{i,N_i}} {\color{black}]}^T$.

\subsection{Feasibility Check (Step 3)}
\label{SubFeasCheck}
A new NCVL is added to {$\bm{\beta}_{i}$} at each iteration of the second step which causes the  number of optimization variables of \eqref{EqOptPrb2} to increase by one and may finally result in the infeasibility of the optimization problem.
As a result, the feasibility of \eqref{EqOptPrb2} should be verified as the third step and after the selection of each NCVL in order to assign the maximum number of unadmitted NCVLs to {$\bm{\beta}_{i}$}. 
Adding unadmitted NCVLs to {$\bm{\beta}_{i}$} should be continued until  \eqref{EqOptPrb2} becomes infeasible, which indicates that the maximum number of NCVLs have been added to $c_{\alpha_i}$.

The optimization problem \eqref{EqOptPrb2} is feasible iff the feasible area {$\mathcal{S}$,} which is the intersection of $(N_{i} + 1)$ half spaces regarding QoS constraints (\ref{EqOpt122} and \ref{EqOpt132}) alongside $(N_{i} + 1)$ half spaces regarding power constraints (\ref{EqOpt152} and \ref{EqOpt162}), is non-empty.
By defining
$g^c_i \triangleq h^c_{\alpha_i}$,
$g^{d,b}_j \triangleq h^{d,b}_{\beta_{i,j}}$,
$g^d_j \triangleq h^d_{\beta_{i,j}}$,
$g^{c,d}_{i,j} \triangleq h^{c,d}_{i,\beta_{i,j}}$, and
$g^{d,d}_{k,j} \triangleq h^{d,d}_{\beta_{i,k},\beta_{i,j}}$
for {notational} simplicity and the fact that the transmission powers are optimization variables,
 the QoS constraints of $c_{\alpha_i}$ and $\beta_{i,j}$ can be expressed as \eqref{EqFeas1} and \eqref{EqFeas3} and then transformed into an affine form as \eqref{EqFeas2} and \eqref{EqFeas4}, respectively, which are expressed in the following:
\begin{equation}
\label{EqFeas1}
\gamma_{\alpha_i}^c = \frac{p^c_{\alpha_i} g^c_i}{\sigma^2
+\sum\limits_{j=1}^{N_i} p^d_{\beta_{i,j}} g^{d,b}_j} \triangleq \frac{A^c_i}{B^c_i} \geq \gamma_{\alpha_i}^{c,\rm{min}},
\end{equation}
\begin{equation}
\label{EqFeas2}
p^c_{\alpha_i} g^c_i  - \sum\limits_{j=1}^{N_i} p^d_{\beta_{i,j}} g^{d,b}_j \gamma_{\alpha_i}^{c,\rm{min}}
\geq \sigma^2 \gamma_{\alpha_i}^{c,\rm{min}},
\end{equation}
\begin{align}
\nonumber
\gamma_{\beta_{i,j}}^d &= \frac{p^d_{\beta_{i,j}} g^d_j}{\sigma^2
+p^c_{\alpha_i} g^{c,d}_{i,j} + \sum\limits_{k=1 \atop k\neq j}^{N_i} p^d_{\beta_{i,k} } g^{d,d}_{k,j}}\\
\label{EqFeas3}
&\triangleq \frac{A^{d}_{i,j}}{B^{d}_{i,j}} \geq \gamma_{\beta_{i,j}}^{d,\rm{min}}, ~~ j=1,...,N_i ,
\end{align}
\begin{align}
\nonumber
-p^c_{\alpha_i} g^{c,d}_{i,j} \gamma_{\beta_{i,j}}^{d,\rm{min}} +p^d_{\beta_{i,j}} g^d_j
- \sum\limits_{k=1 \atop k\neq j}^{N_i} p^d_{\beta_{i,k} } g^{d,d}_{k,j} \gamma_{\beta_{i,j}}^{d,\rm{min}}\\
\label{EqFeas4}
\geq \sigma^2 \gamma_{\beta_{i,j}}^{d,\rm{min}}, ~~ j=1,...,N_i.
\end{align}
Considering \eqref{EqOpt152} and \eqref{EqOpt152}, the power constraints can be expressed as
$0 \leq \bm{{\color{black}p}}_i \leq \bm{{\color{black}p}}_i^{\rm{max}}$,
where
$\bm{{\color{black}p}}_i^{\rm{max}} \triangleq {\color{black}[} P^{c,\rm{max}}_{\alpha_i}, P^{d,\rm{max}}_{\beta_{i,1}} ,..., P^{d,\rm{max}}_{\beta_{i,N_i}} {\color{black}]}^T$
is the maximum transmission power vector of the TX of $c_{\alpha_i}$ and the TXs of NCVLs that reuse $c_{\alpha_i}$.
Using \eqref{EqFeas2} and \eqref{EqFeas4}, the QoS constraints can be expressed as
$\bm{H}_i\bm{{\color{black}p}}_i \geq \sigma^2 \bm{\gamma}_i$,
where the minimum required SINR vector of $c_{\alpha_i}$ and the NCVLs that reuse $c_{\alpha_i}$ can be defined as
$\bm{\gamma}_i \triangleq {\color{black}[} \gamma_{\alpha_i}^{c,\rm{min}}, \gamma_{\beta_{i,1}}^{d,\rm{min}}, ...
\gamma_{\beta_{i,N_i}}^{d,\rm{min}} {\color{black}]}^T$.
{Thus}, the channel gain matrix
$\bm{H}_i \in \mathcal{R}^{(N_i+1)\times (N_i+1)}$ can be formulated as follows:
\begin{subequations}
\label{HDefinition}
\begin{alignat}{2}
&H_{i_{1,1}} && \triangleq g^c_i,\\
&H_{i_{1,j+1}} &&\triangleq -g^{d,b}_j \gamma_{\alpha_i}^{c,\rm{min}},  j=1,...,N_i,\\
&H_{i_{j+1,1}} &&\triangleq -g^{c,d}_{i,j} \gamma_{\beta_{i,j}}^{d,\rm{min}},  j=1,...,N_i,\\
&H_{i_{j+1,k+1}} &&\triangleq -g^{d,d}_{k,j} \gamma_{\beta_{i,j}}^{d,\rm{min}},  j=1,...,N_i,  k=1,...,N_i, k \neq j,\\
&H_{i_{j+1,j+1}} &&\triangleq g^{d}_{j} , ~~~~~~~~~~~~~~ j=1,...,N_i.
\end{alignat}
\end{subequations}

In discussing the third step,
{\color{black}
it should be noted that
one of the main contributions of our proposed scheme is the introduction of the feasibility check theorem (Theorem \ref{FeasCheckTheorem}) which provides necessary and sufficient conditions for the feasibility verification of the optimization problem \ref{EqOptPrb2} before utilizing the optimal power allocation algorithm (Step 4) for solving the optimization problem.
The theorem has not been proposed in the literature previously.
We analytically prove the sufficiency and necessity conditions.
}
We introduce the concepts of $Z$-matrix and $M$-matrix alongside Lemma \ref{Lemma1}, Lemma \ref{Lemma2}, and Lemma \ref{Lemma3} as the concepts and the lemmas are used in the proof of the theorem related to this step.
The matrix $\bm{A} \in \mathcal{R}^{n \times n}$ is a $Z$-matrix of order n (i.e., $\bm{A} \in Z^{n,n}$) if its off-diagonal entries are less than or equal to zero
(i.e., $\bm{A} = (a_{i,j}), ~\mathrm{s.t.} ~ a_{i,j}\leq 0 , \forall i \neq j$)
\cite{R131}.
An M-matrix is a Z-matrix with eigenvalues whose real parts are non-negative \cite{R131}.
\begin{lemma}
\label{Lemma1}
If $\bm{A} \in \mathcal{R}^{n \times n}$ is a Z-matrix and there exist $x>0$ with $\bm{A}x>0$, then $\bm{A}$ is a non-singular M-matrix.
\end{lemma}
~~~\textit{Proof}:
The proof is presented {in} \cite{R131}.
\hfill$\blacksquare$
\begin{lemma}
\label{Lemma2}
If $\bm{A} \in \mathcal{R}^{n \times n}$ is a Z-matrix and a non-singular M-matrix, then $\bm{A}$ is monotone, i.e., if $\bm{A}x \geq 0$, then $x\geq 0$.
\end{lemma}
~~~\textit{Proof}:
The proof is presented {in} \cite{R131}.
\hfill$\blacksquare$
\begin{lemma}
\label{Lemma3}
If the elements of $\bm{A} \in \mathcal{R}^{n \times n}$ are all random numbers, then $|\bm{A}|\neq0$ with probability one.
\end{lemma}
~~~\textit{Proof}:
The proof is presented {in} \cite{S030}.
\hfill$\blacksquare$

\vspace{0.3cm}
The effective Theorem \ref{FeasCheckTheorem} expresses necessary and sufficient conditions for the feasibility check of \eqref{EqOptPrb2}.
The proposed closed-form criterion \eqref{FeasEqTheorem} is the third step of our proposed scheme.
\begin{theorem}[Feasibility Check Theorem]
\label{FeasCheckTheorem}
The feasible area of \eqref{EqOptPrb2} is non-empty iff
\begin{equation}
\label{FeasEqTheorem}
0 \leq \bm{{\color{black}p}}_i^{\textrm{\rm{Init}}} \triangleq \sigma^2 \bm{H}^{-1} \bm{\gamma}_i \leq \bm{{\color{black}p}}_i^{\rm{max}}.
\end{equation}
This means that the point
$\bm{{\color{black}p}}_i^{\textrm{\rm{Init}}} = {\color{black}[} p^{c,\textrm{\rm{Init}}}_{\alpha_i},
p^{d,\textrm{\rm{Init}}}_{\beta_{i,1}},...,
p^{d,\textrm{\rm{Init}}}_{\beta_{i,N_i}} {\color{black}]} $
 is inside the hypercube defined by maximum transmission power parameters; i.e.,
\begin{subequations}
\label{FeasibilityEqs}
\begin{align}
\label{InnerMaxPcCon}
0 \leq &p^{c,\textrm{\rm{Init}}}_{\alpha_i} \leq P^{c,\rm{\rm{max}}}_{\alpha_i}, \\
\label{InnerMaxPdCon}
0 \leq &p^{d,\textrm{\rm{Init}}}_{\beta_{i,j}} \leq P^{d,\rm{\rm{max}}}_{\beta_{i,j}}, ~~ j=1,...,N_i.
\end{align}
\end{subequations}
\end{theorem}
~~~\textit{Proof}:
The proof is presented in Appendix \ref{Appendix1}.
\hfill$\blacksquare$

\subsection{Optimal Power Allocation (Step 4)}
\label{SubPowAlloc}
When no more NCVLs can be added to the admitted NCVLs set of $c_{\alpha_i}$, the optimization problem \eqref{EqOptPrb2} should be solved in order to allocate transmission powers to the TX of $c_{\alpha_i}$ and the TXs of NCVLs that reuse $c_{\alpha_i}$.
The links are selected from the CVL-NCVL matching (Steps 1 and  2) and the feasibility of the optimization problem is verified (Step 3).
By removing the binary resource sharing indicators of  \eqref{OptProbEq} due to the CVL-NCVL matching, the constraints of \eqref{EqOptPrb2} can be expressed in an affine form with respect to $\bm{{\color{black}p}}_i$ as \eqref{EqFeas2} and \eqref{EqFeas4} which causes the feasible area $\mathcal{S}$ to be convex.
The non-convex objective function of \eqref{EqOptPrb2}  can always be expressed as the summation of a concave and a convex function which is called a DC form \cite{R005_021} as follows:
\begin{align}
\nonumber
R^i(\bm{{\color{black}p}}_i)&= \log_2(1+\frac{A^c_i}{B^c_i}) + \sum_{j=1}^{N_i} \log_2(1+\frac{A^{d}_{i,j}}{B^{d}_{i,j}})\\
\label{ObjFuncMaMi}
&= R^{i}_{cav}(\bm{{\color{black}p}}_i) + R^{i}_{vex}(\bm{{\color{black}p}}_i),
\end{align}
where $R^{i}_{cav}(\bm{{\color{black}p}}_i)$ ($R^{i}_{vex}(\bm{{\color{black}p}}_i)$) denotes the summation of strictly concave (convex) positive (negative) logarithm functions that is defined as equation \eqref{FunCav} (\eqref{FunVex}).
Thus, the functions $R^{i}_{cav}(\bm{{\color{black}p}}_i)$ and $R^{i}_{vex}(\bm{{\color{black}p}}_i)$ that are formulated in the following are strictly concave and convex functions, respectively.
\begin{subequations}
\begin{align}
\label{FunCav}
R^{i}_{cav}(\bm{{\color{black}p}}_i) &\triangleq \log_2(B^c_i+A^c_i) + \sum_{j=1}^{N_i} \log_2(B^d_{i,j}+A^d_{i,j}),
\\
\label{FunVex}
R^{i}_{vex}(\bm{{\color{black}p}}_i) &\triangleq - \log_2(B^c_i)  - \sum_{j=1}^{N_i} \log_2(B^d_{i,j}).
\end{align}
\end{subequations}
Due to the DC form of our sum-rate function, our optimization problem can be categorized as a CCCP problem \cite{R005_023}.
The CCCP method investigates the optimization of energy functions with a DC form and states that certain methods can be developed to converge to a minimum or saddle point of the optimization problem.
According to the discussion of \cite{R005_023} and since we aim to maximize the sum-rate of the cell, we conclude that certain algorithms can also be developed for maximizing the DC form sum-rate function.
Hence, by exploiting the DC form of the objective function, we propose an efficient resource allocation algorithm using the MaMi \cite{R005_022} technique and interior point methods.

According to the MaMi method and due to the fact that the optimization problem is a maximization one, we would like to express the objective function $R^i(\bm{{\color{black}p}}_i)$ as a concave function which is also a lower bound for the objective function.
Therefore, we keep the concave term $R^{i}_{cav}(\bm{{\color{black}p}}_i)$ unchanged and approximate the convex term $R^{i}_{vex}(\bm{P}_i)$ using the first order Taylor expansion which is also a lower bound for the term.

Due to the differentiability of  $R^{i}_{vex}(\bm{{\color{black}p}}_i)$, an affine lower bound function $\tilde{R}_{vex}^i (\bm{{\color{black}p}}_i,\bm{{\color{black}p}}_i^{0})$ with respect to $\bm{{\color{black}p}}_i$ can be found for $R^{i}_{vex}(\bm{{\color{black}p}}_i)$ using its first order Taylor expansion around $\bm{{\color{black}p}}_i=\bm{{\color{black}p}}_i^0$ as
\begin{align}
\nonumber
R^{i}_{vex}(\bm{{\color{black}p}}_i) &\geq \nabla R^{i}_{vex}(\bm{{\color{black}p}}^{0}_i)^T (\bm{{\color{black}p}_i} - \bm{{\color{black}p}}^{0}_i) +R^{i}_{vex}(\bm{{\color{black}p}}^{0}_i)\\
\label{EqMinorizorVex}
&\triangleq \tilde{R}_{vex}^i (\bm{{\color{black}p}}_i,\bm{{\color{black}p}}_i^{0}),
\end{align}
where $\nabla R^{i}_{vex}(\bm{{\color{black}p}}^{0}_i)^T$ denotes the transpose of the gradient of $R^i_{vex}(\bm{{\color{black}p}}_i)$ at $\bm{{\color{black}p}}_i = \bm{{\color{black}p}}_i^{0}$.

In the sequel, the MaMi method can be used to solve the optimization problem \eqref{EqOptPrb2}.
According to \cite{S079}, the MaMi method is an iterative technique that can be used to obtain a solution to the general maximization problem of the following form:
\begin{subequations}
\label{MaMiOptProb}
\begin{alignat}{3}
&\max_{\bm{z}} \quad && \tilde{f}(\bm{z}), \\
&\, \rm{s.t.} && c(\bm{z}) \leq 0,
\end{alignat}
\end{subequations}
where $\tilde{f}(.)$ and $c(.)$ are non-convex functions.
Each iteration of the MaMi method consists of a Minorization and a Maximization step.
The Minorization step is to find the minorizor function of $k$th iteration denoted by $\tilde{p}^{(k)}(\bm{z})$ such that
\begin{subequations}
\label{MaMiStepOne}
\begin{align}
\tilde{p}^{(k)}(\bm{z}) &\leq \tilde{f}(\bm{z}), ~ \forall \bm{z},\\
\tilde{p}^{(k)}({\bm{z}}^{(k-1)}) &= \tilde{f}({\bm{z}}^{(k-1)}),
\end{align}
\end{subequations}
where ${\bm{z}}^{(k-1)}$ is the value of $\bm{z}$ at $(k-1)$th iteration.
The Maximization step is to obtain $\bm{z}^{(k)}$ by solving the following optimization problem:
\begin{subequations}
\label{MaMiStepTwo}
\begin{alignat}{3}
&\max_{\bm{z}} \quad && \tilde{p}^{(k)}(\bm{z}), \\
&\, \rm{s.t.} && c(\bm{z}) \leq 0.
\end{alignat}
\end{subequations}

By denoting the $k$th iterating point of the MaMi technique \cite{R005_022} as $\bm{{\color{black}p}}_i^{(k)}$, the technique with the minorizor function $\tilde{R}^i(\bm{{\color{black}p}}_i,\bm{{\color{black}p}}_i^{(k)})$, which is strictly concave with respect to $\bm{{\color{black}p}}_i$, can be used iteratively to solve the optimization problem \eqref{EqOptPrb2} around current iterating point $\bm{{\color{black}p}}_i = \bm{{\color{black}p}}_i^{(k)}$.
The minorizor function can be expressed as {follows:}
\begin{align}
\nonumber
\tilde{R}^i(\bm{{\color{black}p}}_i,\bm{{\color{black}p}}_i^{(k)}) &\triangleq R^i_{cav}(\bm{{\color{black}p}}_i)+\tilde{R}_{vex}^i (\bm{{\color{black}p}}_i,\bm{{\color{black}p}}_i^{(k)})\\
\label{MaMiMinorizor}
&\leq  R^{i}_{vex}(\bm{{\color{black}p}}_i) + R^{i}_{cav}(\bm{{\color{black}p}}_i) = R^i(\bm{{\color{black}p}}_i).
\end{align}
The minorizor is a valid function for the MaMi technique \cite{R005_022} since
\begin{subequations}
\begin{align}
\tilde{R}^i(\bm{{\color{black}p}}_i,\bm{{\color{black}p}}_i^{(k)}) &\leq R^i(\bm{{\color{black}p}}_i), \forall \bm{{\color{black}p}}_i \in \mathcal{S},\\
\tilde{R}^i(\bm{{\color{black}p}}_i^{(k)},\bm{{\color{black}p}}_i^{(k)}) &= R^i(\bm{{\color{black}p}}_i^{(k)}).
\end{align}
\end{subequations}
Hence,
due to the concavity of $\tilde{R}^i(\bm{{\color{black}p}}_i,\bm{{\color{black}p}}_i^{(k)})$ and convexity of $\mathcal{S}$, the optimization problem \eqref{EqOptPrb2} can be solved iteratively from the following convex programming:
\begin{align}
\nonumber
\bm{{\color{black}p}}_i^{(k+1)} &= \arg \max_{\bm{{\color{black}p}}_i \in \mathcal{S}}  \Big\{ \tilde{R}^i(\bm{{\color{black}p}}_i,\bm{{\color{black}p}}_i^{(k)})\Big\} \\
\label{InnerOptMaMi}
&= \arg \max_{\bm{{\color{black}p}}_i \in \mathcal{S}}  \Big\{ R^{i}_{cav}(\bm{{\color{black}p}}_i) + \nabla R^{i}_{vex}(\bm{{\color{black}p}}^{(k)}_i)^T (\bm{{\color{black}p}}_i ) \Big\},
\end{align}
where constant terms are eliminated from the argument maximization formula.
Since the initial non-convex optimization problem \eqref{EqOptPrb2} is transformed into the convex optimization problem \eqref{InnerOptMaMi},
effective methods can be utilized such as interior point methods \cite{ConvexOptBoyd}, which can solve linear convex optimization problems in an effective manner.
Using \eqref{FunVex}, \eqref{EqFeas1}, and \eqref{EqFeas2}, the gradient of $R^{i}_{vex}$ can be computed as follows:
\begin{equation}
\label{GradEq}
\nabla R^{i}_{vex}(\bm{{\color{black}p}}^{(k)}_i) \hspace{-0.1cm} =\hspace{-0.1cm}
\left[\hspace{-0.1cm}
\begin{array}{l}
\sum\limits_{l=1}^{N_i} \frac{(-1) g^{c,d}_{i,l}}{\ln(2) B_{i,l}^{d,(k)}}\\
\frac{(-1)g^{d,b}_j}{\ln(2) B^{c,(k)}_i}
+ \sum\limits_{l=1 \atop l\neq j}^{N_i} \frac{(-1)g^{d,d}_{j,l}}{\ln(2) B^{d,(k)}_{i,l}}, j=1,...,N_i
\end{array}
\hspace{-0.1cm} \right]\hspace{-0.1cm},
\end{equation}
where the
$B^{c,(k)}_i = B^{c}_i \Bigr|_{\bm{{\color{black}p}}_i =\bm{{\color{black}p}}_i^{(k)} } $
and
$B_{i,l}^{d,(k)} = B_{i,l}^{d} \Bigr|_{\bm{{\color{black}p}}_i =\bm{{\color{black}p}}_i^{(k)} } $.
\begin{algorithm}[t]
\caption{Optimal power allocation}
\label{AlgInner}
\begin{algorithmic}[1]
\STATE \textbf{Initialize}:   $k=0, \bm{{\color{black}p}}_i^{(k)}=\bm{{\color{black}p}}_i^{\textrm{Init}}$, { tolerance} $\epsilon \geq 0$
\REPEAT
\STATE Compute $\nabla R^{i}_{vex}(\bm{{\color{black}p}}^{(k)}_i)$ from \eqref{GradEq}
\STATE Compute $\bm{{\color{black}p}}_i^{(k+1)}$ from \eqref{InnerOptMaMi}
\STATE k=k+1
\UNTIL {$||\bm{{\color{black}p}}_i^{(k)} - \bm{{\color{black}p}}_i^{(k-1)}|| \leq \epsilon$}
\end{algorithmic}
\end{algorithm}

The optimal power allocation algorithm for $c_{\alpha_i}$ is expressed in Algorithm \ref{AlgInner}.
{\color{black}
Theorem \ref{TheoremMonInc} describes the monotonic ascent property of the objective function during the optimal power allocation step which is inspired by \cite{R005_024}.
Theorem \ref{CCCPStationaryTh} is related to the convergence of the proposed optimal power allocation scheme to a KKT satisfying stationary point which utilizes Theorem \ref{FeasCheckTheorem} and has not been introduced in the literature previously.
Some concepts from convex optimization topic are used in the proof.
}
\begin{theorem}[Strictly Increasing Behavior]
\label{TheoremMonInc}
If $\bm{{\color{black}p}}_i^{(k+1)} \neq \bm{{\color{black}p}}_i^{(k)}$, then the objective function $R^i(\bm{{\color{black}p}}_i)$ is strictly increasing on the sequence $\bm{{\color{black}p}}_i^{(k)}$ generated by \eqref{InnerOptMaMi}.
\end{theorem}
~~~\textit{Proof}:
The proof is presented in Appendix \ref{Appendix2}.
\hfill$\blacksquare$

\begin{theorem}[Stationary Point Convergence]
\label{CCCPStationaryTh}
If
$0 \leq \bm{{\color{black}p}}_i^{\textrm{\rm{Init}}} \triangleq \sigma^2 \bm{H}^{-1} \bm{\gamma}_i \leq \bm{{\color{black}p}}_i^{\rm{max}}$ ,
then the sequence $\bm{{\color{black}p}}_i^{(k)}$ converges to a point $\bm{{\color{black}p}}_i^{(\infty)}$, which is a stationary point of $R^i(\bm{{\color{black}p}}_i)$ satisfying KKT conditions.
\end{theorem}
~~~\textit{Proof}:
The proof is presented at Appendix \ref{Appendix3}.
\hfill$\blacksquare$

\subsection{Proposed Algorithm}
\label{PrpAlg}
In this section, we propose our overall resource allocation algorithms based on the four previously introduced steps.
Different CVL and NCVL admission orders might be utilized in the CVL-NCVL matching procedure, which would result in algorithmically different overall resource allocation schemes.
{\color{black}
We propose to use the introduced steps but the arrangement of the steps can result in algorithmically different approaches.
}
Considering different sequence of the steps, we propose two different overall resource allocation algorithms denoted as Matching-based Spectrally Efficient Resource Allocation I (MSERA-I) and MSERA-II
{\color{black}
 which are expressed as Algorithm \ref{PrpAlgFirst} and Algorithm \ref{PrpAlgSecond}, respectively
}

\begin{algorithm}[t]
\caption{Matching-based Spectrally Efficient Resource Allocation I (MSERA-I)}
\label{PrpAlgFirst}
\begin{algorithmic}[1]
\STATE \textbf{Initialize}:  $\bm{\beta}_i=\emptyset, \forall i \in \mathcal{C}$  and  $\bm{U}=\mathcal{D}$
\STATE {Compute $\bm{\alpha} = \{ \alpha_1, \alpha_2, ..., \alpha_N \}$ using Algorithm \ref{AlgCUECL}}
\FOR {$i \in {\{1, 2, ..., N\}}$}
\STATE $\bm{U}_i=\bm{U}$ and $K_{feas}=|\bm{U}_i|$
\WHILE {$K_{feas}>0$ }
\STATE Update $B_{feas}, \bm{U}_i,$ and $\bm{U}$ using Algorithm \ref{D2DSelAlg}
\STATE $K_{feas}=|\bm{U}_i|$
\ENDWHILE
\STATE Optimal power allocation for the { UEs reusing} the CL of $c_{\alpha_i}$ using Algorithm \ref{AlgInner}
\ENDFOR
\end{algorithmic}
\end{algorithm}

\begin{algorithm}[t]
\caption{Matching-based Spectrally Efficient Resource Allocation II (MSERA-II)}
\label{PrpAlgSecond}
\begin{algorithmic}[1]
\STATE \textbf{Initialize}:  $\bm{\beta}_i=\emptyset, \forall i \in \mathcal{C}$   and  $\bm{U}=\mathcal{D}$
\STATE {Compute $\bm{\alpha} = \{ \alpha_1, \alpha_2, ..., \alpha_N \}$ using Algorithm \ref{AlgCUECL}}
\STATE $\bm{N}_{feas} = [1, 1, ... ,1]^T \in \{0,1\}^{N\times 1} $
{
\STATE $\bm{U}_i=\bm{U}$
}
\WHILE {$\sum_{i=1}^{N} {\bm{N}_{feas}(i)}>0$ $\&$ $|\bm{U}|>0$ }
\STATE $i=i+1$
{
\STATE $i={\rm mod}(i-1,N)+1$
}
\IF {$\bm{N}_{feas}(i)$}
\STATE $\bm{U}_i = \bm{U}$
\WHILE {$|\bm{U}_i|>0$}
\STATE Update $B_{feas}, \bm{U}_i,$ and $\bm{U}$ using Algorithm \ref{D2DSelAlg}
{
\IF {$|\bm{U}_i|=0$}
\STATE $\bm{N}_{feas}(i)=0$
\ENDIF
\IF {$B_{feas}$}
\STATE {\rm{continue}}
\ENDIF
}
\ENDWHILE
\ENDIF
\ENDWHILE
\STATE Optimal power allocation for all CLs using Algorithm \ref{AlgInner}
\end{algorithmic}
\end{algorithm}

It is necessary to describe the difference among Algorithms MSERA-I and MSERA-II.
Considering the CVL priority and CL assignment which is presented as Algorithm \ref{AlgCUECL}, both algorithms
try to match NCVLs to CVLs with higher priorities before other CVLs.
The MSERA-I algorithm, which is presented as Algorithm \ref{PrpAlgFirst}, performs the matching procedure $N$ times (once for each CVL).
The algorithm tries to match as many unadmitted NCVLs to $c_{\alpha_i}$ as possible by adding the index of newly admitted NCVLs to the index set $\bm{\beta}_i$ as the second step of the proposed scheme.
It should be noted that the index set $\bm{\beta}_i$ corresponds to the CL of the $i$th highest priority CVL.
The feasibility of the resource allocation problem is verified for each CVL-NCVL matching (Step 3) with the fast feasibility check (Theorem \ref{FeasCheckTheorem}), which is presented as Algorithm \ref{D2DSelAlg}.
The matching procedure of each CVL is finalized when no more NCVLs can be{matched with  $c_{\alpha_i}$ while the resource allocation problem remains feasible.
The MSERA-II algorithm, which is described as Algorithm \ref{PrpAlgSecond}, matches one unadmitted NCVL to each CVL (Step 2) at each iteration while the resource allocation problem remains feasible (Step 3).
The feasibility of each matching is verified with Theorem \ref{FeasCheckTheorem}.
The procedure continues until no more NCVLs can be matched to any other CVLs.
The last step is the optimal power allocation (Step 4) with an initial point which was computed as part of Theorem \ref{FeasCheckTheorem} for the CVL-NCVL matching of each CL.

{\color{black}
In other words, both algorithms try to match as many NCVLs to CVLs.
The MSERA-I approach performs this operation in a serial manner where resource sharing is performed on the basis of CVLs.
The MSERA-II approach performs this operation in a semi-parallel manner where resource sharing is performed on the basis of NCVLs.
}

\begin{algorithm}[t]
\caption{NCVL selection and feasibility check}
\label{D2DSelAlg}
\begin{algorithmic}[1]
\STATE NCVL ($d_{new}$) selection using \eqref{SelNextD2DCh}
\STATE {Compute $\bm{H}_i$ using \eqref{HDefinition}}
{
\IF {\eqref{FeasEqTheorem}}
\STATE $B_{feas}=1$
\ELSE
\STATE $B_{feas}=0$
\ENDIF
}
\IF {$B_{feas}$}
\STATE  $k=|\bm{\beta}_{i}|+1$
{
\STATE $\beta_{i,k}=$ $d_{new}$
}
\STATE $\bm{U} = \bm{U} { \textbackslash \{} d_{new} {\}}$
{
\STATE $\bm{U}_i = \bm{U}$
}
\ELSE
\STATE $\bm{U}_i = \bm{U}_i { \textbackslash \{} d_{new}{\}}$
\ENDIF
\end{algorithmic}
\end{algorithm}

\begin{table}[htbp]
\begin{center}
\caption{Simulation Parameters} \label{T2}
\def\arraystretch{1.5}
\small
\begin{tabular}{|p{0.9in}|p{1in}|p{1.5in}|} \hline
\multicolumn{2}{|p{1.5in}|}{\textbf{Parameter}}& \textbf{Value}  \\ \hline
\multicolumn{2}{|p{1.5in}|}{Physical link type} & {Uplink} \\ \hline
\multicolumn{2}{|p{1.5in}|}{Cell radius} & {400 m} \\ \hline
\multicolumn{2}{|p{1.5in}|}{Noise power ($\sigma^2_N$)} & -114 dBm \\ \hline
\multicolumn{2}{|p{1.5in}|}{Path loss model} & $15.3+37.6 \log_{10}D$ (\textit{D} in m) \\ \hline
\multicolumn{2}{|p{1.5in}|}{Max NCVL TX power ($P^{d,\rm{max}}$)} & 21 dBm\\ \hline
\multicolumn{2}{|p{1.5in}|}{Max CVL power ($P^{c,\rm{max}}$)} & 24 dBm\\ \hline
\multicolumn{2}{|p{1.5in}|}{CVL and NCVL min QoS
($\gamma^{c,\rm{min}}$)}
& Uniformly distributed in [0, 10] dB\\ \hline
\multicolumn{2}{|p{1.5in}|}{Shadowing standard deviation} & 8 dB\\ \hline
\multicolumn{2}{|p{1.5in}|}{\vspace{-0.3cm}{Multi-path fading mean}} & {\vspace{-0.3cm}1}\\ \hline
\multicolumn{2}{|p{1.5in}|}{NCVL cluster radius} & Uniformly distributed in [10, 40] m\\ \hline
\multicolumn{2}{|p{1.5in}|}{Number of CVLs (N)} & 6, 8, ..., 16 \\ \hline
\multicolumn{2}{|p{1.5in}|}{Number of NCVLs (M)} & 4N, 6N, ..., 20N \\ \hline
\end{tabular}
\end{center}
\vspace*{-0.6cm}
\end{table}

\section{Numerical Results}
\label{SecNumeric}

{\color{black}
In this section, the performance of the proposed algorithms are evaluated via Monte-Carlo simulations over 1000 channel realization and random user placement.
All simulations are performed on an ordinary PC (with 16GB RAM and CPU CoRe i5).
The MATLAB software is used.
To solve convex optimization problems, we use CVX toolbox in MATLAB with default solver.
}

We consider a single-cell network, where the BS is located at the center of the cell and VUEs are uniformly distributed in the cell.
The network is fully loaded and all CVLs use the total bandwidth of the cell equally.
It can be concluded that the TXs of NCVLs are also uniformly distributed in the cell while RXs of the NCVLs are uniformly distributed in a cluster around their corresponding TX.
The channel gain values are generated by considering a distance-based path loss model, a slow fading gain due to shadowing with log-normal distribution, and a fast fading gain due to multi-path propagation with exponential distribution.
The distance-based path loss model is described in Table \ref{T2}.
Therefore, all channel gain values can be formulated similarly.
Without loss of generality and for instance, $h^c_i$ can be expressed as { follows:}
\begin{equation}
h^c_i = K \zeta^c_i \eta^c_i (L^c_i)^{-\rho},
\end{equation}
where $K$ and $\rho$ { denote} the path loss constant and path loss exponent of the path loss model, respectively.
The distance between the TX of $c_i$ and the BS is denoted by $L^c_i$.
The parameters $\zeta^c_i$ and $\eta^c_i$ denote the slow fading gain with log-normal distribution between the TX of $c_i$ and the BS and the unit mean exponentially distributed fast fading gain between the TX of $c_i$ and the BS, respectively.
The rest of the simulation parameters are summarized in Table \ref{T2}.
The performance evaluation metric is the overall sum-rate of the cell defined as the summation of the spectral efficiencies of all CVLs and admitted NCVLs.

\begin{figure}[htbp]
    \vspace{-0.4cm}
    \hspace{-0.5cm}
    \includegraphics[trim={0cm 0 0cm 0},clip,width=1.2\linewidth]{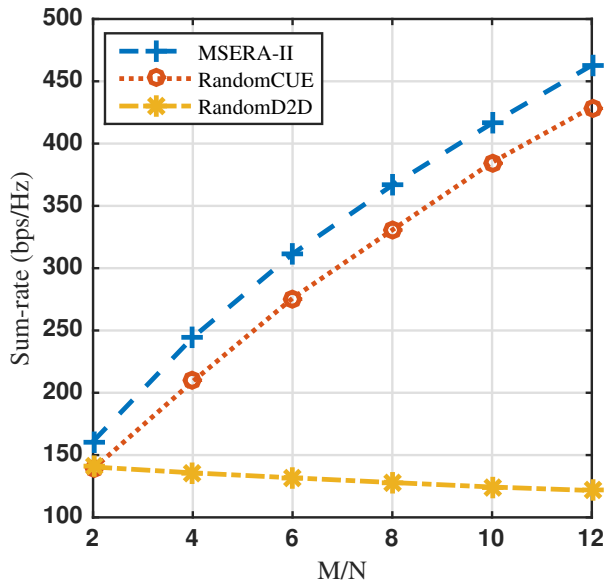}
    \caption{Sum-rate{s} of different CVL priority and CL assignment and NCVL selection methods {relative to} cell density when $N=10$.}
    \label{FigSumRateCDVsMN}
\end{figure}
\begin{figure}[h]
    \vspace{-0.1cm}
    \hspace{-0.5cm}
    \includegraphics[trim={0 0 0cm 0},clip,width=1.2\linewidth]{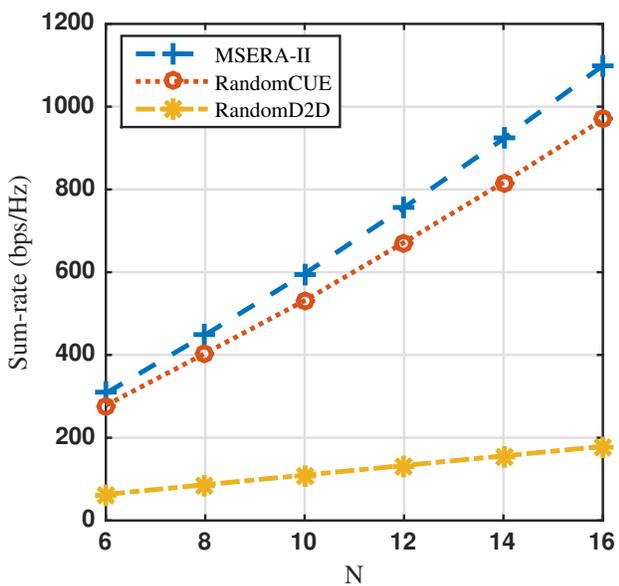}
    \caption{Sum-rate{s} of different CVL priority and CL assignment and NCVL selection methods {relative to} the number of CLs when $M/N=20$.}
    \label{FigSumRateCDVsN}
    \vspace{-0.3cm}
\end{figure}

\subsection{CVL Priority and CL Assignment and NCVL Selection Performance Evaluation}
Figs. \ref{FigSumRateCDVsMN} and \ref{FigSumRateCDVsN} represent the effect of different kinds of CVL-NCVL matchings on the performance of our proposed algorithms.
It should be noted that the MSERA-II algorithm is slightly more complex than the MSERA-I algorithm with higher performance in terms of sum-rate.
Therefore, the MSERA-II algorithm has been selected for the performance comparison figures.
In order to investigate the effect of different CVL-NCVL matchings, MSERA-II, RandomCVL, and RandomNCVL methods are considered.
The MSERA-II method has been illustrated as Algorithm \ref{PrpAlgSecond}.
A channel-gain-based CVL priority and CL assignment, a min-max channel-gain-based NCVL selection procedure, and an optimal power allocation are used in the MSERA-II algorithm.
The RandomCVL and  RandomNCVL methods correspond to the MSERA-II algorithm when random CVL priority and CL assignment is used instead of the proposed CVL priority and CL assignment and random NCVL selection procedure is used instead of the proposed NCVL selection procedure, respectively.

Figs. \ref{FigSumRateCDVsMN} and \ref{FigSumRateCDVsN} represent the sum-rates of different CVL-NCVL matchings for different cell densities and different number of CLs, respectively.
As we can see, the proposed channel-gain-based CVL priority has a better sum-rate performance than the random CVL priority in both figures.
We can also see that the min-max channel-gain-based NCVL selection is an effective method for NCVL selection with a high performance gap compared to the random NCVL pair selection scheme.
The sum-rate performance of the RandomNCVL method decreases by increasing the cell density, since the random NCVL selection becomes less effective when cell density increases.

\subsection{Performance Comparison with Other Methods for Additional CLs}
\begin{figure}[htbp]
    \vspace{-0.2cm}
    \hspace{-0.3cm}
    \includegraphics[trim={0cm 0 0cm 0},clip,width=1.07\linewidth]{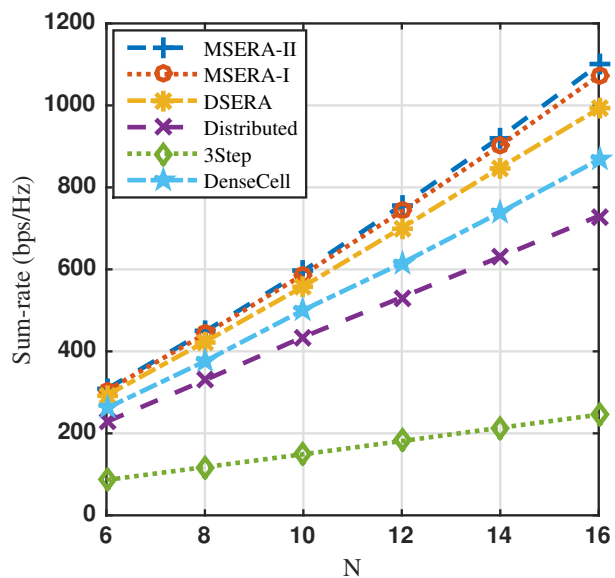}
    \caption{\color{black}Sum-rate{s} of different resource allocation methods {relative to} the number of CLs when $M/N=20$.}
    \label{FigSumRateAllVsN}
    \vspace{-0.3cm}
\end{figure}
\begin{figure}
    \hspace{-0.3cm}
    \includegraphics[trim={0cm 0 0cm 0},clip,width=1.07\linewidth]{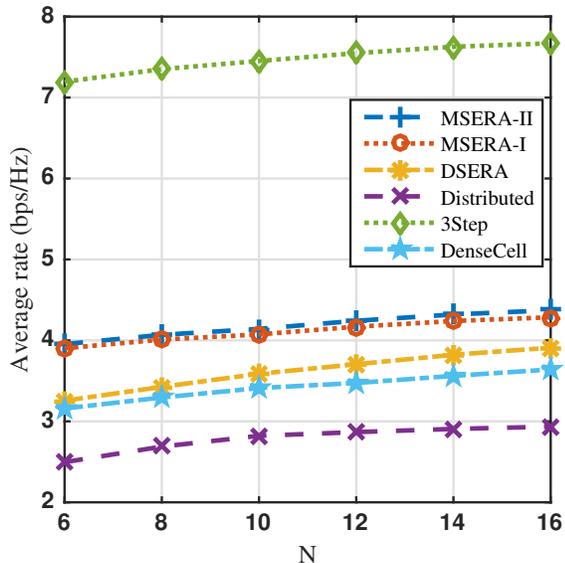}
    \caption{\color{black}Average rate{s} of different resource allocation methods {relative to} the number of CLs when $M/N=20$.}
    \label{FigAvgRateAllVsN}
\end{figure}

We compare the performance of the MSERA-I and MSERA-II methods with that of 3Step, DenseCell, DSERA and Distributed methods, which are slightly modified resource allocation algorithms of \cite{R003}, \cite{R042_026}, \cite{R077_3}, and \cite{R011}, respectively.
In \cite{R003} it is assumed that $M=N$ but due to our $M>N$ assumption, the 3Step method denotes the resource allocation algorithm of \cite{R003} when $M>N$ and Kuhn-Munkres method is applied to all $MN$ possible matchings.
By considering the sum-rate and QoS constraints of CVLs and admitted NCVLs, we define the DenseCell method on the basis of \cite{R042_026} which is a simple and greedy resource allocation algorithm where $M>N$.
The subchannel sharing protocol of \cite{R042_026} evaluates the resource sharing possibility of NCVLs.
The original algorithm of \cite{R011} is an interference coordination algorithm where $M>N$, and it uses a pricing based mechanism to guarantee the QoS requirements of CVLs and NCVLs.
Since the original algorithm does not maximize the sum-rate, we define the Distributed method as an improvement of the original algorithm by assuming that each NCVL reuses the CL of at most one CVL, and by changing the payoff function of NCVLs with a resource sharing indicator as follows:
\begin{equation}
\label{DistributedD2DCost}
U^d_j(\bm{\theta},\bm{P},\bm{\psi})=
p_j^d \sum_{i=1}^N \theta_i \psi_{i,j} \frac{h^{d,b}_j}{h^d_j},
\end{equation}
where $\bm{\theta}=\{ \theta_1, \theta_2, ... , \theta_N\}$ and $\theta_i$ denotes the interference cost of reusing $c_i$.
This improved method uses a NCVL admission mechanism in order to maintain the feasibility of the resource allocation problem for dense C-V2X communications which, maximizes the sum-rate of the cell in a distributed manner.

Figs. \ref{FigSumRateAllVsN} and \ref{FigAvgRateAllVsN} offer a comparison of the sum-rates and average data rates of all methods relative to $N$, respectively.
Fig. \ref{FigSumRateAllVsN} indicates that our proposed methods outperform other methods in terms of sum-rate for additional CLs when the cell is dense.
The performance gap between the MSERA-I and MSERA-II methods increases by increasing $N$.
The sum-rate of the 3Step method is much lower than the other methods since at most $N$ NCVLs are admitted.
The average data rate of each method is shown in Fig. \ref{FigAvgRateAllVsN}.
{\color{black}
It should be mentioned that the average rate is the ratio of the sum-rate to the the number of VCLs and admitted NCVLs since VCLs and NCVLs are considered in the sum-rate formula \eqref{Eq_Rt}.
}
As we can see, the average data rate of our method is higher than that of the Distributed and DenseCell methods but lower than that of the 3Step method, since multiple NCVLs are able to reuse each CL in all of the methods except the 3Step method.
Therefore, the average data rate of the 3Step method is much higher than that of the others.
It should be noted that our optimization problem and proposed method are based on maximizing the sum-rate of the cell, and therefore the average data rate is not a performance evaluation metric.

\subsection{Performance Comparison with Other Methods for Different Cell Densities}
\begin{figure}[htbp]
    \hspace{-0.3cm}
    \includegraphics[trim={0cm 0 0cm 0},clip,width=1.07\linewidth]{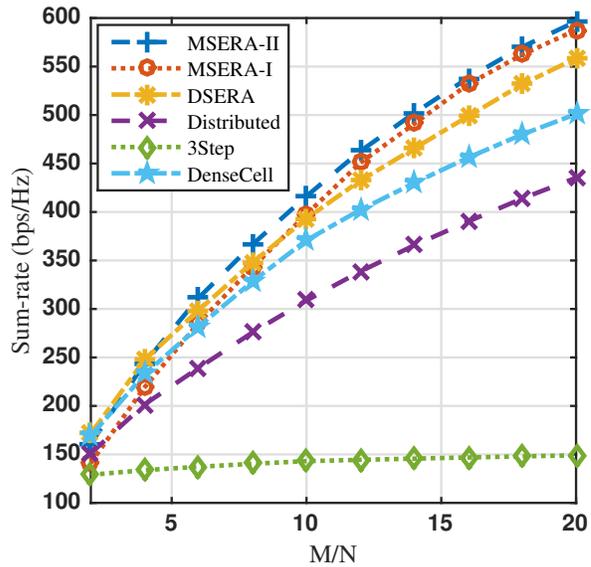}
    \caption{\color{black}Sum-rate{s} of different resource allocation methods relative to cell density when $N=10$.}
    \label{FigSumRateAllVsMN}
\end{figure}
\begin{figure}
    \hspace{-0.3cm}
    \includegraphics[trim={0cm 0 0cm 0},clip,width=1.07\linewidth]{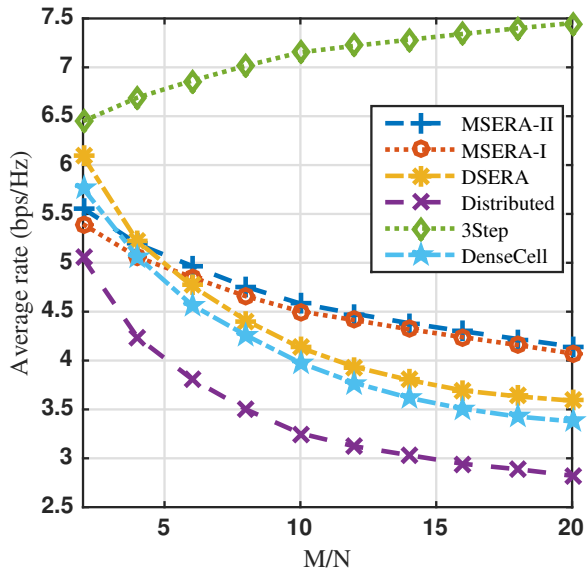}
    \caption{\color{black}Average rate{s} of different resource allocation methods relative to cell density when $N=10$.}
    \label{FigAvgRateAllVsMN}
\end{figure}

Figs. \ref{FigSumRateAllVsMN} and \ref{FigAvgRateAllVsMN} present comparisons between the sum-rates and average data rates relative to $M/N$, designating cell densities, respectively.
As we can see in Fig. \ref{FigSumRateAllVsMN}, the performance gap between the MSERA-I and MSERA-II methods decreases when the cell density increases.
When the value of $M/N$ is low, which means that the cell is not dense, the sum-rate performance of the DenseCell method is greater than that of our proposed methods due to the former's higher computational complexity and full CSI requirement.
By increasing the cell density parameter ($M/N$), our proposed methods outperform other methods in terms of sum-rate due to our proposed simple, innovative, and scalable CVL-NCVL matching procedure combined with the optimal power allocation.
{\color{black}
The average data rate of the 3Step method is higher than that of the other methods for dense scenarios since at most one NCVL can reuse each CL.
The average data rate of the 3Step method is also increasing since each CVL would be able to share its resource with a larger number of NCVLs when the cell density is increased.
Hence, a better NCVL in terms of sum-rate would be chosen by the modified Hungarian algorithm for each CVL when the cell density is increased for a fixed number of CVLs.
}
As we can see, the Distributed and DenseCell methods are not effective schemes for maximizing the spectral efficiency of dense C-V2X communications.
Our proposed scheme is effective for such dense scenarios in next generation cellular networks with numerous transmitting entities.

{\color{black}
Cell density is an important and challenging aspect of future wireless networks.
Cell density change affects all layers of the proposed scheme and it is a challenging test for the feasibility of the proposed scheme.
We have investigated the sum-rate of the proposed scheme at different levels of cell density which is represented by the $M/N$ factor .
By investigating the performance of our proposed scheme for low, high, very-high, and ultra-high cell densities and
according to Figs. \ref{FigSumRateAllVsMN} and \ref{FigAvgRateAllVsMN}, it is observed that the proposed scheme is feasible for mild changes of the cell density parameter that affects all layers of the proposed scheme.
Hence, the proposed scheme if a feasible solution.
}

\subsection{Comparison of MSERA-I and MSERA-II}
The MSERA-I method has a slightly less complex implementation than the MSERA-II method with lower sum-rate performance as we can see in Fig. \ref{FigSumRateAllVsN}.
Since both algorithms have the same performance when the cell density increases according to Fig. \ref{FigSumRateAllVsMN}, the MSERA-I and MSERA-II methods are suggested for low and high-density scenarios, respectively.
The MSERA-I algorithm has a serial matching approach while the approach of the MSERA-II algorithm is semi-parallel.

\subsection{Computational Complexity}
The proposed algorithms are both low complexity sub-optimal approaches since not all possible CVL-NCVL matchings are checked, but an innovative procedure is used to find  sub-optimal matchings.
The run-times of different algorithms are presented in Fig. \ref{PicRunTime}.
Our proposed method has a shorter run-time than the DenseCell algorithm but a longer run-time than the 3Step and Distributed algorithms.
However, the overall sum-rate of the proposed algorithm is significantly greater than that of the 3Step and Distributed algorithms in dense scenarios.
The 3Step algorithm is fast since the algorithm allows each CL to be used by at most one NCVL.
{\color{black}
The run-time complexity of MSERA-I and MSERA-II are also shown in Fig. \ref{PicRunTime}.
It is observed that both algorithm have almost the same complexity while MSERA-II is slightly more complex due to its more complex algorithmic implementation.
}

{\color{black}
Due to the almost similar complexity of MSERA-I and MSERA-II algorithms, we consider MSERA-I for analytic complexity expression.
There are $N$ iteration corresponding to each CVL.
Each CVL shares its resource with $M/N$ NCVLs on average.
The evaluation of the Algorithm \ref{D2DSelAlg} is also density dependent.
Hence, the analytical complexity expression would be $\mathcal{O}(N (\frac{M}{N})^{\alpha})$ where $\alpha=2.1$ is estimated from Fig. \ref{PicRunTime}.
The derived expression is implementation and platform based similar to many existing algorithm.
For instance, sparse processing algorithms can be utilized to reduce the complexity due to the spare structure of variables.
}

\begin{figure}[t]
    \hspace{-0.2cm}
    \includegraphics[trim={0cm 0 0cm 0},clip,width=1.12\linewidth]{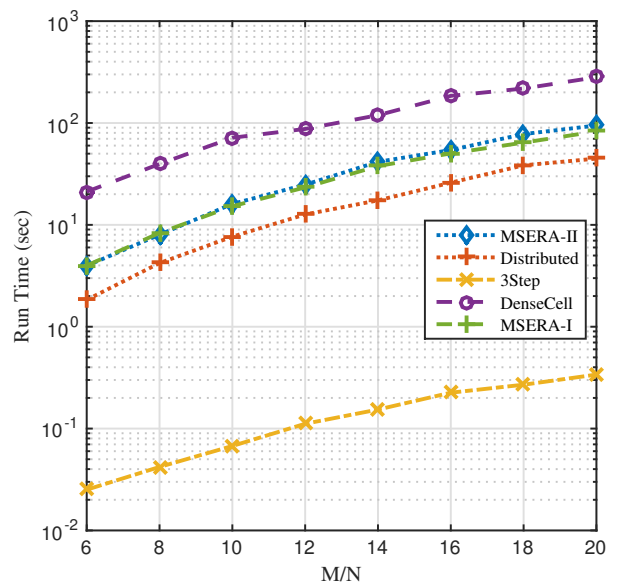}
    \vspace{-0.5cm}
    \caption{\color{black}The comparison of the run-time of our proposed method and other resource allocation algorithms when $N=10$.}
    \vspace{-0.5cm}
    \label{PicRunTime}
\end{figure}

{\color{black}
\subsection{Convergence}
The proposed optimization method (Step 4) allocates optimal powers to admitted UEs.
The convergence of the optimization method for sum-rate maximization is shown in Fig. \ref{PicMaMiConv} via considering the values of objective function \eqref{ObjFuncMaMi} versus iteration numbers.
The figure verifies that the objective function values have a monotonic ascent property, as expected and analytically proved in Theorem \ref{TheoremMonInc}.
\begin{figure}[t]
    \hspace{-0.6cm}
    \includegraphics[trim={4.5cm 0 4.3cm 0},clip,width=1.2\linewidth]{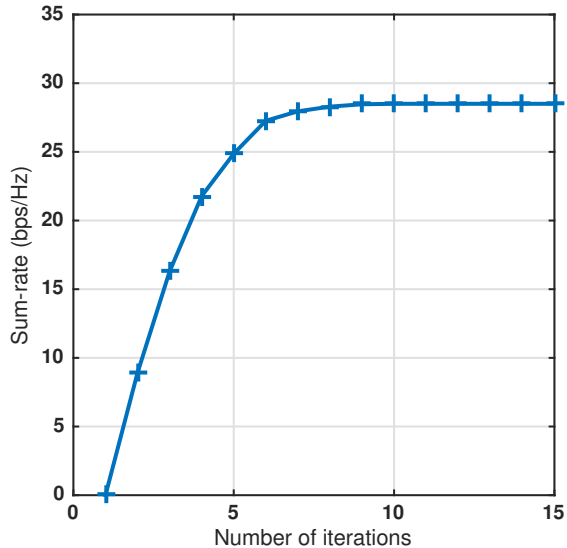}
    \vspace{-0.5cm}
    \caption{{\color{black}The value of objective function versus number of iterations.}}
    \vspace{-0.5cm}
    \label{PicMaMiConv}
\end{figure}

}

\subsection{Upper bound and Optimality Gap}
Due to the extremely high number of possible matchings related to the optimal scheme, it is impossible to compute the optimal sum-rate of the cell for all cases.
It should be noted that the optimal approach is to compute sum-rates of all possible matchings using the optimal power allocation step (Step 4) followed by the Kuhn-Munkres algorithm to find the optimal CVL-NCVL matching.
Instead of the optimal sum-rate, an upper bound derived by solving the relaxation of the overall resource allocation problem \eqref{OptProbEq} using the \texttt{fmincon} solver can be used.
When $N=10$ and the cell is ultra-dense, the \texttt{fmincon} solver cannot solve the relaxed form of \eqref{OptProbEq}.
As a result, we investigate the upper bound when $N=3$ and $M/N$ varies from 2 to 16.
Fig. \ref{PicOptGapLim} presents a comparison of the sum-rate performance of different resource allocation methods with the derived upper bound.
\begin{figure}[t]
    \hspace{0.5cm}
    \includegraphics[trim={0cm 0 0cm 0},clip,width=0.95\linewidth]{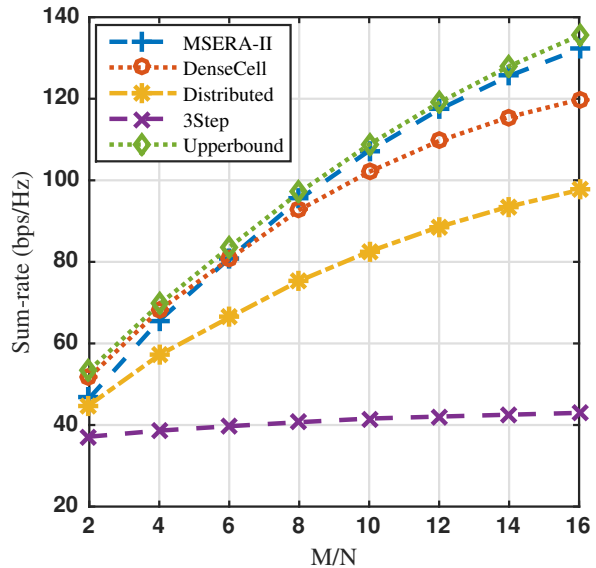}
    \vspace{-0.5cm}
    \caption{\color{black}Sum-rate comparison of different resource allocation methods with an upper bound when $N=3$.}
    \label{PicOptGapLim}
\end{figure}
As we can see, the performance of our proposed scheme is lower than that of the DenseCell method in a low-density scenario due to the former's simple and effective CVL-NCVL matching and its lower computational complexity.
Most existing methods are not scalable for dense C-V2X communications.
Hence, the optimality gap of other methods increases when the cell density increases.
For instance, the DenseCell method investigates the resource sharing possibility of each two NCVLs but does not investigate the resource sharing possibility for a greater number of NCVLs.
Therefore, the DenseCell method has a high sum-rate performance when $M/N=2$ but its optimality gap increases when the cell density increases.
Our proposed scheme has a simple and scalable CVL-NCVL matching followed by an optimal power allocation which results in a low optimality gap in dense C-V2X communications.
The optimality gap of our proposed method decreases when the cell density increases until a certain cell density value is reached.
The optimality gap remains acceptable but slightly increases when the cell density increases further.
We conclude that our proposed scheme is close to optimum in dense C-V2X communications.

\subsection{Signaling Overhead}
The channel gain of all desired and interference links among TXs of CVLs and BS, TXs of CVLs and RXs of NCVLs, and TXs of NCVLs and RXs of NCVLs are required for the optimal scheme, which is therefore a method with a full CSI requirement and large signaling overhead.
The DenseCell method is also a method with full CSI requirement due to its sub-channel sharing protocol.
The channel gain of all links is not required in the Distributed method, which is therefore a method with a partial CSI requirement but needs to exchange interference prices among UEs, resulting in signaling overhead increment.
The 3Step method does not need channel gains between TXs of NCVLs and RXs of NCVLs, which therefore makes it a method with a partial CSI requirement.
According to the algorithmic representation of our proposed scheme described in Algorithm \ref{PrpAlgFirst} and Algorithm \ref{PrpAlgSecond}, a newly admitted NCVL is removed from the set of unadmitted NCVLs.
The admitted NCVL is selected using the NCVL selection procedure (Step 2) and verified using the feasibility check theorem (Step 3).
Not all channel gain values are required in the NCVL selection procedure
since the channel gain values among the TX of the newly admitted NCVL and the RXs of previously admitted NCVLs that use the same CL and among the RX of the newly admitted NCVL and the TXs of previously admitted NCVLs that use the same CL are required.
Thus, the step is a partial CSI requirement step.
Since other steps of our proposed algorithm are also partial CSI requirement steps, our proposed method is a partial CSI requirement method where not all channel gains are required.

{\color{black}
\subsection{Mobility}
In order to verify the effectiveness of the proposed scheme under mobility assumption, we develop a simulation setup where multiple VUEs with straight line mobility model exist in the cell.
As the VUE moves, the VUE is in need of link establishment with new nodes in its vicinity and should terminate the links that are not required any more.
It is assumed that $N$ mobile VUEs exist in the cell.
Each VUE establishes one CVL and an average number of $M/N$ NCVLs (new NCVLs may be established and some of old NVCLs may be terminated).
It is also assumed that the mobile VUEs move in straight lines according to the straight line mobility model from the start time ($t=0$) to the final time ($t=T_{max}$).
In order to evaluate the performance of our proposed scheme for the mentioned mobility model, we evaluate the sum-rate of our proposed scheme at times $t=k/L T_{max},~ k=0, 1, ..., (L-1)$.
Fig. \ref{PicMobility} demonstrates the sum-rate performance of the network for $N=3$ and $L=10$ versus time with different cell densities.
It is shown that the sum-rate performance is guaranteed and its value is consistent with other simulation results.
Therefore, is can be concluded that the proposed scheme is effective under the mobile VUE assumption.
}

\begin{figure}[t]
    \hspace{0.5cm}
    \includegraphics[trim={0cm 0 0cm 0},clip,width=0.95\linewidth]{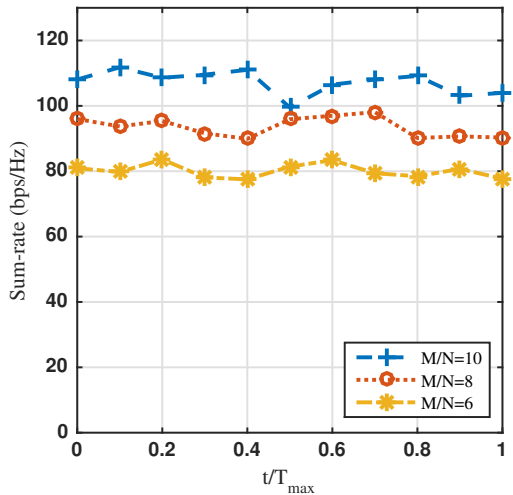}
    \vspace{-0.5cm}
    \caption{\color{black}Summ-rate versus time for VUEs with straight mobility assumption when $N=3$.}
    \label{PicMobility}
\end{figure}

\section{Conclusion}
\label{Conc}
In this paper, we investigated centralized resource allocation for dense C-V2X communications in future cellular networks.
Our approach was based on leveraging the spatial reuse gain of cellular networks to admit as many NCVLs as possible.
The first two steps of our approach resulted in a channel-gain-based CVL-NCVL matching.
The effectiveness of the proposed channel-gain-based CVL priority and CL assignment and NCVL selection in terms of spectral efficiency was then evaluated numerically.
We also discussed the scalability of our proposed resource allocation methods.
It was shown that the low optimality gap due to the scalability property is acceptable.
The scalability property and the optimal power allocation result in a low optimality gap, which indicates our proposed resource allocation algorithms will be practical in dense future wireless networks with numerous transmitting entities.
The signaling overhead of our proposed methods was shown to be lower than that of the optimal approach since the proposed resource allocation algorithms can be implemented using partial CSI.
We also discussed the computational complexity of the proposed approach and justified its practicality on the basis of the simple, fast, and analytically proven feasibility check theorem.
The numerical results demonstrated that the performance of the proposed scheme is acceptable but lower than that of other competitive methods in low-density scenarios due to the lower computational complexity of the former.
Our results also confirmed that the performance of our proposed algorithms is higher than that of other methods for dense and ultra-dense C-V2X communications.
Hence, our proposed MSERA-I and MSERA-II algorithms are low-complexity, effective, practical, and scalable for dense C-V2X communications in next generation cellular networks.
{\color{black}
Resource allocation to dense mobile VUEs underlaying cellular network via optimization approaches can be a future research direction.
}

\appendices

\section{Proof of theorem 1}
\label{Appendix1}
\textit{Proof of sufficiency}: It is known that $\bm{{\color{black}p}}_i^{\textrm{Init}} = \sigma^2 \bm{H}^{-1} \bm{\gamma}_i$ and hence
$\bm{H}_i \bm{{\color{black}p}}_i^{\textrm{Init}} = \sigma^2 \bm{\gamma}_i \geq \sigma^2 \bm{\gamma}_i$ which means that QoS constraints hold.
Additionally, $0 \leq \bm{{\color{black}p}}_i^{\textrm{Init}} \leq \bm{{\color{black}p}}^{\rm{max}}_i$ which means that power constraints hold.
So, the point  $\bm{{\color{black}p}}_i^{\textrm{Init}}$ is a feasible point of \eqref{EqOptPrb2} since both QoS and power constraints hold. Hence, the feasible area is not empty i.e., $\mathcal{S} \neq \emptyset$.

\noindent
\textit{Proof of necessity}: Since the feasible area is not empty,
$\exists \tilde{\bm{{\color{black}p}}}\geq 0$, s.t. $\bm{H}_i \tilde{\bm{{\color{black}p}}} \geq \sigma^2 \bm{\gamma}_i > 0$.
Since $\bm{H}_i$ is a $Z$-matrix and $\exists \tilde{\bm{{\color{black}p}}} \geq 0$ with $\bm{H}_i \tilde{\bm{{\color{black}p}}} \geq 0$, then $\bm{H}_i$ is a non-singular M-matrix according to Lemma \ref{Lemma1} and hence is monotone according to Lemma \ref{Lemma2}, i.e.,
if $\bm{H}_i \bm{{\color{black}p}}_i \geq 0$, then $\bm{P}_i \geq 0$.
Due to non-singularity of $\bm{H}_i$ and according to Lemma \ref{Lemma3}, $\bm{H}_i^{-1}$ and $\bm{{\color{black}p}}_i^{\textrm{Init}}$ exist.
From the monotone property of Lemma \ref{Lemma2} and
$\bm{H}_i \bm{{\color{black}p}}_i^{\textrm{Init}} = \sigma^2 \bm{\gamma}_i \geq 0$, {we can conclude} that $\bm{{\color{black}p}}_i^{\textrm{Init}} \geq 0$.

In order to prove $\bm{{\color{black}p}}_i^{\textrm{Init}} \leq \bm{{\color{black}p}}^{\rm{max}}_i$,
{ let us} assume that $\bm{{\color{black}p}}_i^{\textrm{Init}}$ is not a feasible point by assuming
$\bm{{\color{black}p}}_i^{\textrm{Init}} > \bm{{\color{black}p}}^{\rm{max}}_i$.
Since the feasible area is not empty,
$\exists \bar{\bm{{\color{black}p}}}\geq 0$ such that
$\bm{H}_i \bar{\bm{{\color{black}p}}} \geq \sigma^2  \bm{\gamma}_i$, which results in
$\bm{H}_i( \bar{\bm{{\color{black}p}}}-\bm{{\color{black}p}}_i^{\textrm{Init}} ) \geq \sigma^2  \bm{\gamma}_i - \bm{H}_i\bm{{\color{black}p}}_i^{\textrm{Init}}=0$.
According to Lemma \ref{Lemma2} and considering the monotone property of $\bm{H}_i$, it can be concluded that
$\bar{\bm{{\color{black}p}}}-\bm{{\color{black}p}}_i^{\textrm{Init}} \geq 0$ and hence
$\bar{\bm{{\color{black}p}}} \geq \bm{{\color{black}p}}_i^{\rm{Init}}$.
So, $\bar{\bm{{\color{black}p}}} \geq \bm{{\color{black}p}}_i^{\textrm{Init}} > \bm{{\color{black}p}}^{\rm{max}}_i$,
which means $\bar{\bm{{\color{black}p}}}$ is an infeasbile point that is a contradiction and the proof is complete.

\section{Proof of theorem2}
\label{Appendix2}
Inspired by \cite{R005_024} and
due to { the strict} convexity of $R_{vex}^i(\bm{{\color{black}p}}_i)$ and affinity of $\tilde{R}_{vex}(\bm{{\color{black}p}}_i,\bm{{\color{black}p}}_i^{(k)})$, \eqref{EqMinorizorVex} and \eqref{MaMiMinorizor} hold with strict inequality
when $\bm{{\color{black}p}}_i \neq \bm{{\color{black}p}}_i^{(k)}$.
Thus,
\begin{equation}
\label{EqProofOne}
R^i(\bm{{\color{black}p}}_i^{(k+1)}) > \tilde{R}^i(\bm{{\color{black}p}}_i^{(k+1)},\bm{{\color{black}p}}_i^{(k)}),
\end{equation}
when $\bm{{\color{black}p}}_i^{(k+1)} \neq \bm{{\color{black}p}}_i^{(k)}$.
According to \eqref{InnerOptMaMi} and due to { the strict} convexity of $R_{vex}^i(\bm{{\color{black}p}}_i)$, { we can conclude} that
\begin{equation}
\label{EqProofTwo}
\tilde{R}^i(\bm{{\color{black}p}}_i^{(k+1)},\bm{{\color{black}p}}_i^{(k)}) > \tilde{R}^i(\bm{{\color{black}p}}_i^{(k)},\bm{{\color{black}p}}_i^{(k)})=R^i(\bm{{\color{black}p}}_i^{(k)}),
\end{equation}
when $\bm{{\color{black}p}}_i^{(k+1)} \neq \bm{{\color{black}p}}_i^{(k)}$.
Considering  \eqref{EqProofOne} and \eqref{EqProofTwo}, { we can conclude } that
$R^i(\bm{{\color{black}p}}_i^{(k+1)}) > R^i(\bm{{\color{black}p}}_i^{(k)})$ when $\bm{{\color{black}p}}_i^{(k+1)} \neq \bm{{\color{black}p}}_i^{(k)}$ and the proof is complete.

\section{Proof of theorem 3}
\label{Appendix3}
According to the { Theorem} \ref{FeasCheckTheorem}, if $0 \leq \bm{{\color{black}p}}_i^{\textrm{Init}} \triangleq \sigma^2 \bm{H}^{-1} \bm{\gamma}_i \leq \bm{{\color{black}p}}_i^{\rm{max}}$ then $\mathcal{S} \neq \emptyset$.
Considering the randomness of coefficients of \eqref{EqOpt122} and \eqref{EqOpt132} which are expressed in an affine form in \eqref{EqFeas2} and \eqref{EqFeas4}, their corresponding half-spaces are not parallel to the half-spaces corresponding to \eqref{EqOpt152} and \eqref{EqOpt162}.
Hence, $\mathcal{S}$ which is the intersection of the half-spaces is closed and bounded.
Due to the strictly increasing behavior of $R^i(\bm{{\color{black}p}}_i^{(k)})${, it} can be concluded that $\{ \bm{{\color{black}p}}_i^{(k)} \}$ converges to the limit point $\bm{{\color{black}p}}_i^{(\infty)}$ as
$\lim_{k \rightarrow \infty} ||\bm{{\color{black}p}}_i^{(k+1)} - \bm{{\color{black}p}}_i^{(k)}|| = 0 $.
Due to the linearity of constraints \eqref{EqFeas2}, \eqref{EqFeas4}, \eqref{EqOpt152}, and \eqref{EqOpt162} , the Salter's constraint qualification is reduced to feasibility \cite{ConvexOptBoyd}.
Thus, strong duality holds and KKT conditions are satisfied for \eqref{InnerOptMaMi}.
KKT conditions also hold for \eqref{EqOptPrb2} at $\bm{{\color{black}p}}_i^{(\infty)}$ and  the point is a stationary point of \eqref{EqOptPrb2} and the proof is complete.

\bibliographystyle{IEEEtran}
\bibliography{Test}

\begin{IEEEbiography}[{\includegraphics[width=1in,height=1.25in,clip,keepaspectratio]{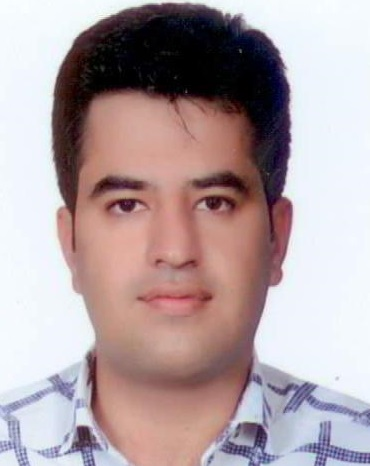}}]{Mohammad Hossein Bahonar}
received the B.Sc. and M.Sc. degrees in electrical engineering from the University of Tehran (UT), Iran, and the Sharif University of Technology (SUT), Iran, in 2012 and 2014, respectively.
He is currently pursuing the Ph.D. degree with the Isfahan University of Technology (IUT), Iran.
He was a visiting researcher with the Politecnico di Torino, Italy, in 2020.
His research interests include cellular networks, resource management, device-to-device communications, and wireless communications.
\end{IEEEbiography}

\begin{IEEEbiography}[{\includegraphics[width=1in,height=1.25in,clip,keepaspectratio]{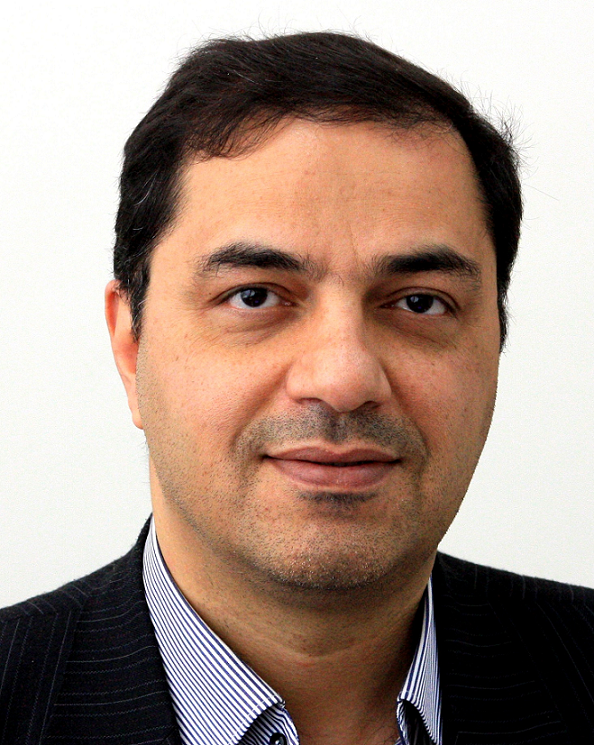}}]{Mohammad Javad Omidi}
received the Ph.D. degree from the University of Toronto in 1998. He was with the industry by joining the Research and Development Group designing broadband communication systems for five years in U.S. and Canada. In 2003, he joined the Department of Electrical and Computer Engineering, Isfahan University of Technology (IUT), Iran, and then served as the Chair of Information Technology Center and the Chair of the ECE Department. He is currently the Director of IRIS a UNESCO organization in Iran, and the Vice President for Research and Development with Isfahan Science and Technology Town. He is currently the supervising the Software Radio Laboratory, IUT and his scientific research interests are in the areas of mobile computing, wireless communications, digital communication systems, software radio, cognitive radio, and VLSI architectures for communication algorithms. He has authored over 15 U.S. and International patents in the area of telecommunications.
\end{IEEEbiography}

\begin{IEEEbiography}[{\includegraphics[width=1in,height=1.25in,clip,keepaspectratio]{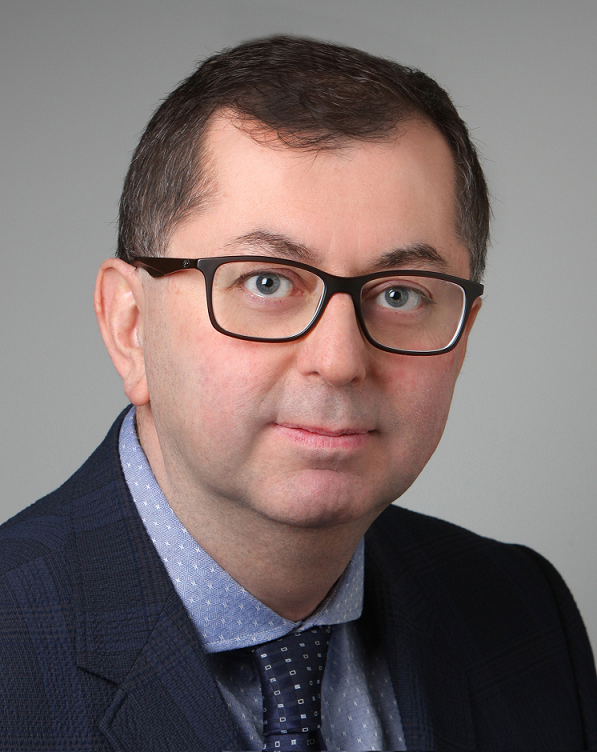}}]{Halim Yanikomeroglu}
is a Professor in the Department of Systems and Computer Engineering at Carleton University, Ottawa, Canada. His primary research domain is wireless communications and networks. His research group has made substantial contributions to 4G and 5G wireless technologies. His group’s current focus is the aerial (UAV and HAPS) and satellite networks for the 6G and beyond-6G era. His extensive collaboration with industry resulted in 37 granted patents. He is a Fellow of IEEE, EIC (Engineering Institute of Canada), and CAE (Canadian Academy of Engineering), and a Distinguished Speaker for both IEEE Communications Society and IEEE Vehicular Technology Society. He is currently serving as the Chair of the IEEE WCNC (Wireless Communications and Networking Conference) Steering Committee. He served as the General Chair or TP Chair of several conferences including three WCNCs and two VTCs. He also served as the Chair of the IEEE’s Technical Committee on Personal Communications. Dr. Yanikomeroglu received several awards for his research, teaching, and service, including the IEEE ComSoc Fred W. Ellersick Prize in 2021, IEEE VTS Stuart Meyer Memorial Award in 2020, and IEEE ComSoc Wireless Communications Technical Committee Recognition Award in 2018.
\end{IEEEbiography}

\end{document}